\DeclareFontFamily{OT1}{msb}{}{}
\DeclareFontShape{OT1}{msb}{m}{n}
 {  <5> <6> <7> <8> <9> <10> gen * msbm
      <10.95><12><14.4><17.28><20.74><24.88>msbm10}{}
\DeclareMathAlphabet{\bubble}{OT1}{msb}{m}{n}

\def\bN{{\bubble N}}
\def\bZ{{\bubble Z}}

\newfont{\bbd}{msbm10 scaled\magstep1}

\documentstyle[12pt]{article}

\parskip=\medskipamount
\begin{document}

\def\l#1#2{\raisebox{.0ex}{$\displaystyle
  \mathop{#1}^{{\scriptstyle #2}\rightarrow}$}}
\def\r#1#2{\raisebox{.0ex}{$\displaystyle
\mathop{#1}^{\leftarrow {\scriptstyle #2}}$}}

\newcommand{\p}[1]{(\ref{#1})}

\newcommand{\sect}[1]{\setcounter{equation}{0}\section{#1}}

\makeatletter
\def\eqnarray{\stepcounter{equation}\let\@currentlabel=\theequation
\global\@eqnswtrue
\global\@eqcnt\z@\tabskip\@centering\let\\=\@eqncr
$$\halign to \displaywidth\bgroup\@eqnsel\hskip\@centering
  $\displaystyle\tabskip\z@{##}$&\global\@eqcnt\@ne
  \hfil$\displaystyle{\hbox{}##\hbox{}}$\hfil
  &\global\@eqcnt\tw@ $\displaystyle\tabskip\z@
  {##}$\hfil\tabskip\@centering&\llap{##}\tabskip\z@\cr}
\makeatother


\renewcommand{\thefootnote}{\fnsymbol{footnote}}
\newpage
\setcounter{page}{0}
\pagestyle{empty}
\begin{flushright}
{November 2000}\\
{JINR E2-270-2000}\\
{nlin.SI/0011009}
\end{flushright}
\vfill

\begin{center}
{\LARGE {\bf Supersymmetric Toda lattice}}\\[0.3cm]
{\LARGE {\bf hierarchies\footnote{To be published in Proceedings 
of the NATO ARW ``Integrable Hierarchies and Modern Physical
Theories'' (Chicago, USA, July 22 - 26, 2000), Kluwer Academic 
Publishers.}}}\\[1cm]

{\large V.G. Kadyshevsky$^{a}$ and A.S. Sorin$^{b}$}
{}~\\
\quad \\

{{Bogoliubov Laboratory of Theoretical Physics,}}\\
{{Joint Institute for Nuclear Research,}}\\
{\em 141980 Dubna, Moscow Region, Russia}~\quad\\
\em {~$~^{(a)}$ Email: kadyshev@jinr.dubna.su}\\
\em {~$~^{(b)}$ Email: sorin@thsun1.jinr.ru}\\

\end{center}

\vfill

\centerline{{\bf Abstract}}
\noindent
The origin of the bosonic and fermionic solutions, constructed in
\cite{ly1,ls1,ols1}, to the symmetry equations corresponding to the
two-dimensional bosonic and $N=(2|2)$ supersymmetric Toda lattices is
established, and algebras of the corresponding symmetries are derived.
The conjecture regarding an $N=(2|2)$ superfield formulation of the
$N=(2|2)$ supersymmetric Toda lattice hierarchy, proposed in \cite{ols2},
is proved. The two-dimensional $N=(0|2)$ supersymmetric Toda lattice
hierarchy is proposed and its $N=(0|2)$ superfield formulation is discussed.
Bosonic and fermionic solutions to the symmetry equation corresponding to the
two-dimensional $N=(0|2)$ supersymmetric Toda lattice equation and their
algebra are constructed. An infinite class of new two-dimensional
supersymmetric Toda-type hierarchies is discussed.

{}~



\vfill
\newpage
\pagestyle{plain}
\renewcommand{\thefootnote}{\arabic{footnote}}
\setcounter{footnote}{0}

\section{Introduction}
Recently, an infinite class of solutions to the symmetry equation of the
two-dimensional Toda lattice (2DTL) has been described in \cite{ly1} in
the framework of a rather heuristic algorithm of simple calculations proposed
there. This algorithm resembles a computer program: It is necessary to
perform many identical operations that can be interrupted at an arbitrary
step and thus obtain relevant information about a system of (2+1)-dimensional
evolution equations belonging to the integrable 2DTL hierarchy. Then, this
algorithm has been generalized to the case of the $N=(1|1)$ supersymmetric
2DTL equation, and an infinite class of bosonic solutions to its symmetry
equation has been constructed in \cite{ls1}. However, supersymmetry suggests
that the symmetry equation possesses fermionic solutions as well, and they
are responsible for fermionic flows of the hierarchy. Ref. \cite{ols1}
dwelled upon this problem and derived a wide class of fermionic solutions.
Bosonic and fermionic solutions generate bosonic and fermionic flows of the
$N=(1|1)$ supersymmetric 2DTL hierarchy in the same way as their bosonic
counterparts -- the solutions to the symmetry equation of the 2DTL -- produce
flows of the 2DTL hierarchy.

For a more complete understanding of an equation and its solutions it seems
necessary to know as many solutions to the corresponding symmetry equation
as possible. But a symmetry equation represents a complicated nonlinear
functional equation, and its both general and particular solutions are not
known in general. Moreover, in general there even exists no algorithm
to solve this problem. As an illustration of the latter fact, let us
mention, e.g. the unsolved yet problem of constructing symmetries to the
$N=(0|2)$ supersymmetric 2DTL equation proposed in \cite{lds}. Due to this
reason the algorithm developed in \cite{ly1,ls1,ols1} and the resulting
solutions to the symmetry equations corresponding to the 2DTL and $N=(1|1)$
2DTL equations, as they have been presented so far, may appear to have come
out of the blue, and it is interesting to understand their origin. In this
connection, let us point out that long ago one of the authors of the
present paper (V.G.K.), when analyzing an application of difference equations
to solving problems of mathematical physics, developed an efficient approach
to constructing solutions of some relativistic equations (for details, see
refs. \cite{vgk1,vgk2,vgk3,vgk4,vgk5,vgk6} and references therein). Then
this approach was successfully applied in \cite{vgk7,vgk8} to
investigate the field theory with the momentum space of a constant curvature
where difference equations naturally arise in the configuration space and the
lattice spacing is defined by the inverse radius of the curvature of the
initial momentum space. Furthermore, it was recently adapted in \cite{vgk9}
to the case of the gauge field theory. It turns out that this approach is
also instructive in the context of the problem under consideration. Thus, our
goal here is to establish the origin of the algorithm and symmetries of refs.
\cite{ly1,ls1,ols1} by reproducing them in the framework of the previously
known integrable discrete hierarchies -- the 2DTL \cite{ut} and super-Toda
lattice (STL) \cite{i} hierarchies -- containing the 2DTL and $N=(1|1)$ 2DTL
equations, respectively, as subsystems.

It is time to explain how we were led to this construction. Refs.
\cite{ly1,ls1,ols1,vgk1,vgk2,vgk3,vgk4,vgk5,vgk6} and \cite{ut,i} can be
considered ancestors of the present paper. As one might
suspect, there is a correspondence between symmetries of the 2DTL and
$N=(1|1)$ 2DTL equations and flows of the latter hierarchies, but this
correspondence is however rather nontrivial. Thus, the 2DTL (STL) hierarchy
has been defined in \cite{ut} (\cite{i}) as a system of infinitely many
equations for {\it infinitely many} fields, while the 2DTL ($N=(1|1)$ 2DTL)
equation involves only a {\it single independent}
(super)field $v_{0,j}$. From the point of view of the former approach the
derivation of symmetries of
the 2DTL ($N=(1|1)$ 2DTL) equation corresponds to extracting those 2DTL (STL)
hierarchy equations which can be realized in terms of the (super)field
$v_{0,j}$ alone after excluding all other (super)fields of the 2DTL (STL)
hierarchy. Keeping in mind this correspondence it is quite natural to suppose
that the algorithm \cite{ly1,ls1,ols1} of constructing symmetries of the 2DTL
and $N=(1|1)$ 2DTL equations is encoded in the structure of these
hierarchies. In the present paper, we demonstrate by explicit construction
that this is indeed the case at least with respect to bosonic symmetries.

The paper is organized as follows. In sections 2 and 3, we reproduce the
algorithm \cite{ly1,ls1,ols1} of constructing bosonic symmetries of the 2DTL
and $N=(1|1)$ 2DTL equations starting with the 2DTL \cite{ut} and STL
\cite{i} hierarchies, respectively, following the methodology of
operating with difference equations developed in
\cite{vgk1,vgk2,vgk3,vgk4,vgk5,vgk6}. Furthermore, we also establish
algebras of both fermionic and bosonic symmetries which were only conjectured
in \cite{ly1,ls1,ols1}, discuss peculiarities of constructing fermionic
symmetries as well as propose related new problems to be solved in future. In
section 3, as a byproduct we also prove the proposed in \cite{ols2} conjecture
regarding an $N=(2|2)$ superfield formulation of the STL hierarchy. In section
4, we solve the problem of constructing solutions to the symmetry equation
corresponding to the $N=(0|2)$ supersymmetric 2DTL. Thus, we first propose
the new $N=(0|2)$ supersymmetric 2DTL hierarchy which contains the $N=(0|2)$
2DTL equation and then construct both bosonic and fermionic symmetries of the
latter equation as well as their algebra. We also discuss an $N=(0|2)$
superfield formulation of the $N=(0|2)$ 2DTL hierarchy. Section 5 is devoted
to a generalization: We propose a wide class of new supersymmetric
integrable hierarchies whose first representative is the $N=(0|2)$ 2DTL
hierarchy. In section 6, we present a short summary of the main results
obtained in the paper.

\section{2DTL hierarchy and symmetries of 2DTL equation}

In this section, based on the 2DTL hierarchy of ref. \cite{ut}, we develop a
general scheme for constructing symmetries of the 2DTL equation and their
algebra and as a byproduct establish the origin of the symmetries constructed
in \cite{ly1}.

\subsection{Lax pair representation and flows}

Our starting point is the Lax pair representation of the
2DTL hierarchy \cite{ut,ks}:
\begin{eqnarray}
\partial^{\pm}_n L^{+}&=&[((L^{\pm})^{n})_{\pm}~,~ L^{+}], \nonumber\\
\partial^{\pm}_n L^{-}&=&[((L^{\pm})^{n})_{\pm}~,~ L^{-}], \quad
n \in {\bN},
\label{laxrepr1}
\end{eqnarray}
with the two Lax operators $L^{+}$ and $L^{-}$,
\begin{eqnarray}
L^{+}=\sum^{\infty}_{k=0} u_{k,j}e^{(1-k){\partial}}, \quad
L^{-}=\sum^{\infty}_{k=0} v_{k,j}e^{(k-1){\partial}},
\label{lax1}
\end{eqnarray}
\begin{eqnarray}
u_{0,j}\equiv 1, \quad v_{0,j}\neq 0,
\label{bound1}
\end{eqnarray}
which generates the abelian algebra of the flows
\begin{eqnarray}
[\partial^{\pm}_n~,~\partial^{\pm}_l]=
[\partial^{+}_n~,~\partial^{-}_l]=0.
\label{algebra1}
\end{eqnarray}
Here, the bosonic fields $u_{k,j}\equiv u_{k,j}(\{t^{+}_n,t^{-}_n\})$
and $v_{k,j}\equiv v_{k,j}(\{t^{+}_n,t^{-}_n\})$ are
defined on the lattice, $j \in {\bZ}$, and $t^{\pm}_n$ are evolution times;
$\partial^{\pm}_n:={\frac{\partial}{\partial t^{\pm}_n}}$ and
the subscript $+$ ($-$) means the part of an operator which includes
operators $e^{l{\partial}}$ at $l\geq 0$ ($l< 0$). Hereafter, the operator
$e^{l{\partial}}$ ($l \in {\bZ}$) is the discrete lattice shift which acts
according to the rule
\begin{eqnarray}
e^{l{\partial}} u_{k,j} \equiv u_{k,j+l} e^{l{\partial}}, \quad
e^{l{\partial}} v_{k,j} \equiv v_{k,j+l} e^{l{\partial}}.
\label{rule1}
\end{eqnarray}
In this section, we will also use the following useful notation
\begin{eqnarray}
(L^{+})^{m}:=\sum^{\infty}_{k=0} u^{(m)}_{k,j}e^{(m-k){\partial}}, \quad
(L^{-})^{m}:=\sum^{\infty}_{k=0} v^{(m)}_{k,j}e^{(k-m){\partial}},
\label{lax2}
\end{eqnarray}
where $\{u^{(m)}_{k,j},~v^{(m)}_{k,j}\}$ ($u^{(1)}_{k,j}\equiv
u_{k,j},~v^{(1)}_{k,j}\equiv v_{k,j}$) are the functionals of the original
fields $\{u_{k,j},~v_{k,j}\}$
whose explicit form is not important for further consideration but
the explicit form of the following functionals:
\begin{eqnarray}
u^{(m)}_{0,j}=1
\label{imp1}
\end{eqnarray}
which can easily be found using eqs. \p{bound1}.

The following set of operator equations:
\begin{eqnarray}
\partial^{\pm}_n(L^{+})^{m}&=&[((L^{\pm})^{n})_{\pm}~,~(L^{+})^{m}],
\nonumber\\
\partial^{\pm}_n (L^{-})^{m}&=&[((L^{\pm})^{n})_{\pm}~,~
(L^{-})^{m}], \quad n,m \in {\bN}
\label{laxrepr2}
\end{eqnarray}
is identically satisfied as a consequence of eqs. \p{laxrepr1}, and the
corresponding system of evolution equations for the functionals
$\{u^{(m)}_{k,j},~v^{(m)}_{k,j}\}$ can easily be derived from them.
It reads
\begin{eqnarray}
\partial^{+}_n u^{(m)}_{k,j}=\sum^{n}_{p=0}
(u^{(n)}_{p,j}u^{(m)}_{k-p+n,j-p+n} -
u^{(n)}_{p,j-k+p-n+m}u^{(m)}_{k-p+n,j}),
\label{equations1}
\end{eqnarray}
\begin{eqnarray}
\partial^{-}_n u^{(m)}_{k,j}=\sum^{n-1}_{p=0}
(v^{(n)}_{p,j}u^{(m)}_{k+p-n,j+p-n} -
v^{(n)}_{p,j-k-p+n+m}u^{(m)}_{k+p-n,j}),
\label{equations2}
\end{eqnarray}
\begin{eqnarray}
\partial^{+}_n v^{(m)}_{k,j}=\sum^{n}_{p=0}
(u^{(n)}_{p,j}v^{(m)}_{k+p-n,j-p+n} -
u^{(n)}_{p,j+k+p-n-m}v^{(m)}_{k+p-n,j}),
\label{equations3}
\end{eqnarray}
\begin{eqnarray}
\partial^{-}_n v^{(m)}_{k,j}=\sum^{n-1}_{p=0}
(v^{(n)}_{p,j}v^{(m)}_{k-p+n,j+p-n} -
v^{(n)}_{p,j+k-p+n-m}v^{(m)}_{k-p+n,j}),
\label{equations4}
\end{eqnarray}
where all fields $\{u^{(m)}_{k,j},~v^{(m)}_{k,j}\}$
in the right-hand side should be put equal to zero at $k < 0$.

\subsection{Symmetries of 2DTL equation}

The 2DTL hierarchy (\ref{laxrepr1}--\ref{lax1}) is
a system of infinitely many equations
for {\it infinitely many\/} fields $\{u_{k,j},~v_{k,j}\}$,
while the 2DTL equation
\begin{eqnarray}
\partial^{-}_1\partial^{+}_1\ln v_{0,j}= -v_{0,j+1} +2 v_{0,j} - v_{0,j-1}
\label{todab}
\end{eqnarray}
represents its first flow and involves only the {\it single \/} lattice field
$v_{0,j}$ and two evolution derivatives, $\partial^{-}_1$ and
$\partial^{+}_1$. The 2DTL equation \p{todab} can be read from eqs.
(\ref{equations2}--\ref{equations3}) if eqs. \p{equations2} are restricted
to the values $\{n=m=k=1\}$
\begin{eqnarray}
\partial^{-}_1 u_{1,j} = v_{0,j} - v_{0,j+1}
\label{todab1}
\end{eqnarray}
and eqs. \p{equations3} to $\{n=m=1, k=0\}$
\begin{eqnarray}
\partial^{+}_1 v_{0,j}= v_{0,j}(u_{1,j} - u_{1,j-1}),
\label{todab2}
\end{eqnarray}
and then the field $u_{1,j}$ is eliminated from eqs.
(\ref{todab1}--\ref{todab2}).

Now we would like to demonstrate that symmetries of the 2DTL equation
\p{todab} can be decoded from the system (\ref{equations1}--\ref{equations4}).

First, one can easily observe the existence of the following subalgebra
of the flow algebra \p{algebra1}:
\begin{eqnarray}
[\partial^{+}_1~,~\partial^{\pm}_n]=
[\partial^{-}_1~,~\partial^{\pm}_n]=0
\label{algebra2}
\end{eqnarray}
which is valid by construction of the 2DTL hierarchy, i.e. the flows
$\partial^{\pm}_n$ of the 2DTL hierarchy commute simultaneously with both
the derivatives ($\partial^{+}_1$ and $\partial^{-}_1$) entering into the
2DTL equation \p{todab}. The latter remarkable property is a necessary, but
not sufficient condition for the flows $\partial^{\pm}_n$ to form symmetries
of the 2DTL equation \p{todab}. Nevertheless, keeping in mind this property
it is quite natural to suppose that the flows $\partial^{\pm}_n$ form
the symmetries (see, ref. \cite{ks} for slightly different arguments),
although to complete the proof, we have additionally to show that they can
in fact be realized in terms of the 2DTL field $v_{0,j}$ alone.

Second, with the last goal in mind let us discuss the above-mentioned
candidates to be the symmetries,
\begin{eqnarray}
\partial^{+}_n v_{0,j} = + v_{0,j}(u^{(n)}_{n,j}-u^{(n)}_{n,j-1})
\label{flows+}
\end{eqnarray}
and
\begin{eqnarray}
\partial^{-}_n v_{0,j} = -v_{0,j}(v^{(n)}_{n,j}-v^{(n)}_{n,j-1}),
\label{flows-}
\end{eqnarray}
in more detail. When deriving these equations,
we have used eqs. (\ref{equations3}--\ref{equations4}) at $\{m=1, k=0\}$ and
the useful relation
\begin{eqnarray}
\sum^{n}_{p=0}(v^{(n)}_{p,j}v_{n-p,j+p-n} -
v^{(n)}_{p,j+n-p-1}v_{n-p,j})=0
\label{identity}
\end{eqnarray}
resulting from the simple, obvious identitity $(L^-)^nL^{-}-L^{-}(L^-)^n=0$.
It turns out that the functionals $u^{(n)}_{n,j}$ and $v^{(n)}_{n,j}$ in the
right-hand side of eqs. (\ref{flows+}) and (\ref{flows-}) can in fact be
expressed in terms of the field $v_{0,j}$ alone by excluding all other fields
of the 2DTL hierarchy by means of the flows $\partial^{-}_1$ and
$\partial^{+}_1$, respectively, entering into the system
(\ref{equations1}--\ref{equations4}). In order to see that, let us analyze
eqs. \p{equations2} and \p{equations3}, respectively, at $n=1$
\begin{eqnarray}
\partial^{-}_1 u^{(m)}_{k,j}=
v_{0,j}u^{(m)}_{k-1,j-1} -
v_{0,j-k+m+1}u^{(m)}_{k-1,j},
\label{rec1}
\end{eqnarray}
\begin{eqnarray}
\partial^{+}_1 v^{(m)}_{k,j} - v^{(m)}_{k,j} (u_{1,j} - u_{1,j+k-m})=
v^{(m)}_{k-1,j+1}-v^{(m)}_{k-1,j}.
\label{rec2}
\end{eqnarray}
Equation \p{rec2} can easily be transformed into a more useful form
for a further analysis which is similar to eq.  \p{rec1}.
Thus, using the relations
\begin{eqnarray}
u_{1,j} - u_{1,j+k-m}=
+\partial^{+}_1 \ln \prod^{m-k}_{n=1}v_{0,j+k-m+n}, \quad m>k
\label{todab2from1}
\end{eqnarray}
resulting from eq. \p{todab2} and introducing the new basis
$v^{(m)}_{k,j}$ $\Rightarrow$ ${\widetilde v}^{(m)}_{k,j}$, according to the
formulae
\begin{eqnarray}
v^{(m)}_{m,j}={\widetilde v}^{(m)}_{m,j}, \quad
v^{(m)}_{k,j}={\widetilde v}^{(m)}_{k,j}\prod^{m-k}_{n=1}v_{0,j+k-m+n}
\quad m>k,
\label{basis1}
\end{eqnarray}
eq. \p{rec2} becomes
\begin{eqnarray}
\partial^{+}_1 {\widetilde v}^{(m)}_{k,j}=
v_{0,j+1}{\widetilde v}^{(m)}_{k-1,j+1} -
v_{0,j+k-m}{\widetilde v}^{(m)}_{k-1,j}, \quad m\geq k.
\label{rec2basis1}
\end{eqnarray}
A simple inspection of \p{rec1} and \p{rec2basis1} shows
that they in fact allow one to express $u^{(n)}_{n,j}$ and
$v^{(n)}_{n,j}$ in terms of $v_{0,j}$. Indeed, eq. \p{rec1}
( \p{rec2basis1} ) represents a recurrent relation connecting the functional
$u^{(n)}_{k,j}$ (${\widetilde v}^{(n)}_{k,j}$) with $u^{(n)}_{k-1,i}$
(${\widetilde v}^{(n)}_{k-1,i}$). Being iterated with the simple starting
value $u^{(n)}_{0,j}=1$ \p{imp1} (${\widetilde v}^{(n)}_{0,j}=1$), it
generates a very nontrivial expression for the functional $u^{(n)}_{n,j}$
($v^{(n)}_{n,j} \equiv {\widetilde v}^{(n)}_{n,j}$) in terms of $v_{0,j}$
after the $n$-th step of the iteration procedure.
The latter, in turn, yield the symmetries $\partial^{+}_n v_{0,j}$ and
$\partial^{-}_n v_{0,j}$ to the 2DTL equation \p{todab} via eqs. \p{flows+}
and \p{flows-}.

Let us remark that the 2DTL equation \p{todab} possesses the
involution
\begin{eqnarray}
(\partial^{\pm}_1)^{\star}=\partial^{\mp}_1, \quad
(v_{0,j})^{\star}= v_{0,i-j},
\label{autob1}
\end{eqnarray}
which relates the symmetries \p{flows+}, \p{rec1} with the symmetries
\p{flows-}, (\ref{basis1}--\ref{rec2basis1}), according to the following
rule:
\begin{eqnarray}
(\partial^{\pm}_n)^{\star}=\partial^{\mp}_n, \quad
(u^{(m)}_{k,j})^{\star}= {\widetilde v}^{(m)}_{k,i-j-1}, \quad
({\widetilde v}^{(m)}_{k,j})^{\star}=u^{(m)}_{k,i-j-1},
\label{autob2}
\end{eqnarray}
where $i\in {\bZ}$ is a fixed number. Besides the involution
(\ref{autob1}--\ref{autob2}), there exists also another involution
\begin{eqnarray}
(\partial^{\pm}_n)^{\bullet}=\partial^{\pm}_n, \quad
(v_{0,j})^{\bullet}= v_{0,i-j}.
\label{autob3}
\end{eqnarray}
Applying this involution to equations \p{flows-} and \p{rec2basis1}
and introducing the following notation:
\begin{eqnarray}
({\widetilde v}^{(m)}_{k,j})^{\bullet}:= u^{(m)-}_{k,i-j-1}, \quad
u^{(m)}_{k,j}:= u^{(m)+}_{k,j},
\label{autob4}
\end{eqnarray}
equations (\ref{flows+}--\ref{flows-}), \p{rec1} and \p{rec2basis1}
can be rewritten in the following unified form:
\begin{eqnarray}
\partial^{\pm}_n v_{0,j}&=& v_{0,j}(u^{(n)\pm}_{n,j}-u^{(n)\pm}_{n,j-1}),
\nonumber\\ \partial^{\mp}_1 u^{(n)\pm}_{k,j}&=&
v_{0,j}u^{(n)\pm}_{k-1,j-1} -
v_{0,j-k+n+1}u^{(n)\pm}_{k-1,j}, \quad u^{(n)\pm}_{0,j}=1.
\label{flows+-}
\end{eqnarray}

The symmetries \p{flows+-} of the 2DTL equation \p{todab} reproduce the
solutions to the corresponding symmetry equation derived first in \cite{ly1}
by a rather heuristic construction, while the algebra \p{algebra1} of the
symmetries was not established there. One of the advantages of the
general, algorithmic procedure, developed here, is the derivation of both the
symmetries \p{flows+-} and their algebra \p{algebra1}. As a byproduct we have
also established the origin of the symmetries discussed in \cite{ly1}.

In next sections we extend the above-developed scheme of
constructing symmetries of the 2DTL equation to the case of supersymmetric
Toda lattices and discuss supersymmetric peculiarities related to fermionic
flows and their algebras.

\section{N=(2$|$2) supersymmetric 2DTL hierarchy}

In this section, we establish the relationship between bosonic symmetries of
the $N=(1|1)$ supersymmetric 2DTL equation constructed in \cite{ls1} and
bosonic flows of the STL hierarchy of ref. \cite{i}. We also discuss the
subtle point for a relation between fermionic symmetries of the $N=(1|1)$
2DTL equation constructed in \cite{ols1} and fermionic flows of the STL
hierarchy \cite{i}, and establish the algebra of both bosonic and fermionic
symmetries.

\subsection{Lax pair representation and flows}

The Lax pair representation of the STL hierarchy is \cite{i}
\begin{eqnarray}
D^{\pm}_n L^{+}&=&(-1)^n{(((L^{\pm})^{n}_{*})}_{\pm})^{*} L^{+}-
(L^{+})^{*(n)}((L^{\pm})^{n}_{*})_{\pm} \nonumber\\
&+&\frac{1}{2}(1\pm 1) (1-(-1)^n)(L^{+})^{n+1}_{*},
\nonumber\\
D^{\pm}_n L^{-}&=&(-1)^n{(((L^{\pm})^{n}_{*})}_{\pm})^{*} L^{-}-
(L^{-})^{*(n)}((L^{\pm})^{n}_{*})_{\pm} \nonumber\\
&+&\frac{1}{2}(1\mp 1) (1-(-1)^n)(L^{-})^{n+1}_{*}, \quad n \in {\bN},
\label{laxreprs1}
\end{eqnarray}
\begin{eqnarray}
L^{+}=\sum^{\infty}_{k=0} u_{k,j}e^{(1-k){\partial}}, \quad
L^{-}=\sum^{\infty}_{k=0} v_{k,j}e^{(k-1){\partial}},
\label{laxs1}
\end{eqnarray}
\begin{eqnarray}
u_{0,j}\equiv 1, \quad v_{0,j}\neq 0,
\label{bound2}
\end{eqnarray}
and it generates the non-abelian algebra of the flows
\begin{eqnarray}
[D^{+}_{n}~,~D^{-}_{l}\}=[D^{\pm}_{n}~,~D^{\pm}_{2l}]=0,\quad
\{D^{\pm}_{2n+1}~,~D^{\pm}_{2l+1}\}=2D^{\pm}_{2(n+l+1)}
\label{algebras1}
\end{eqnarray}
which may be realized via
\begin{eqnarray}
D^{\pm}_{2n} ={\partial}^{\pm}_{2n}, \quad
D^{\pm}_{2n+1} ={\partial}^{\pm}_{2n+1}+
\sum^{\infty}_{l=1}t^{\pm}_{2l-1}{\partial}^{\pm}_{2(k+l)},
\label{covder}
\end{eqnarray}
where $D^{\pm}_{2n}$ and $t^{\pm}_{2n}$ ($D^{\pm}_{2n+1}$ and
$t^{\pm}_{2n+1}$) are bosonic (fermionic) evolution derivatives
and times, respectively; $u_{2k,j}(\{t^{+}_n,t^{-}_n\})$ and
$v_{2k,j}(\{t^{+}_n,t^{-}_n\})$ \\ ( $u_{2k+1,j}(\{t^{+}_n,t^{-}_n\})$
and $v_{2k+1,j}(\{t^{+}_n,t^{-}_n\})$ ) are bosonic (fermionic) lattice
fields. Hereafter, the subscripts (superscripts) $*$ ( $*{(n)}$ and $*$ )
are defined according to the rule \cite{i}:
\begin{eqnarray}
&&(L^{\pm})^{2n}_{*}:=((L^{\pm})^{*}L^{\pm})^{n}, \quad
(L^{\pm})^{2n+1}_{*}:=L^{\pm}((L^{\pm})^{*}L^{\pm})^{n}, \nonumber\\
&&(L^{\pm})^{*(2n)}:=L^{\pm}, \quad
(L^{\pm})^{*(2n+1)}:=(L^{\pm})^{*}, \nonumber\\
&& (L^{\pm}[u_{k,j},v_{k,j}])^{*}:=L^{\pm}[u_{k,j}^{*},v_{k,j}^{*}],
\quad (u_{k,j},v_{k,j})^{*}:=(-1)^k (u_{k,j},v_{k,j}).
\label{rule2}
\end{eqnarray}

In this section we will use the following notation:
\begin{eqnarray}
(L^{+})^{m}_{*}:=\sum^{\infty}_{k=0} u^{(m)}_{k,j}e^{(m-k){\partial}}, \quad
(L^{-})^{m}_{*}:=\sum^{\infty}_{k=0} v^{(m)}_{k,j}e^{(k-m){\partial}},
\label{laxs2}
\end{eqnarray}
where $\{u^{(m)}_{2k,j},~v^{(m)}_{2k,j}\}$  and
$\{u^{(m)}_{2k+1,j},~v^{(m)}_{2k+1,j}\}$ ($u^{(1)}_{k,j}\equiv
u_{k,j},~v^{(1)}_{k,j}\equiv v_{k,j}$) are bosonic and fermionic functionals
of the original fields $\{u_{k,j},~v_{k,j}\}$ whose explicit form is not
important in what follows but
\begin{eqnarray}
u^{(m)}_{0,j}=1.
\label{imp2}
\end{eqnarray}

The operator equations
\begin{eqnarray}
D^{\pm}_n (L^{+})^{m}_{*}&=&
(-1)^{nm}(((L^{\pm})^{n}_{*})_{\pm})^{*(m)} (L^{+})^{m}_{*}
-((L^{+})^{m}_{*})^{*(n)}((L^{\pm})^{n}_{*})_{\pm}\nonumber\\
&+&\frac{1}{2}(1\pm 1) (1-(-1)^{nm})(L^{+})^{n+m}_{*}, \nonumber\\
D^{\pm}_n (L^{-})^{m}_{*}&=&
(-1)^{nm}(((L^{\pm})^{n}_{*})_{\pm})^{*(m)} (L^{-})^{m}_{*}-
((L^{-})^{m}_{*})^{*(n)}((L^{\pm})^{n}_{*})_{\pm} \nonumber\\
&+&\frac{1}{2}(1\mp 1) (1-(-1)^{nm})(L^{-})^{n+m}_{*}, \quad n,m \in {\bN}
\label{laxreprs2}
\end{eqnarray}
are identically satisfied on the shell of the original equations
\p{laxreprs1} and reproduce the latter at the value $m=1$. Let us remark that
they can identically be rewritten in a rather standard Lax pair form if
artificial fermionic parameters ${\epsilon}_n$ and ${\varepsilon}_m$ are
introduced,
\begin{eqnarray}
&&\Bigl({\prod}^{n}_{k=1}{\epsilon}_k\Bigr)
D^{\pm}_n {\prod}^{m}_{p=1}({\varepsilon}_pL^{\mp})=
\Bigl[\Bigl({\prod}^{n}_{k=1}({\epsilon}_kL^{\pm})\Bigr)_{\pm}~,~
{\prod}^{m}_{p=1}({\varepsilon}_pL^{\mp})\Bigr], \nonumber\\
&&\Bigl({\prod}^{n}_{k=1}{\epsilon}_k\Bigr)
D^{\pm}_{n} {\prod}^{m}_{p=1}({\varepsilon}_pL^{\pm})=
(-1)^{nm}\nonumber\\
&&\times\Bigl[\Bigl({\prod}^{n}_{k=1}
({\epsilon}_kL^{\pm})\Bigr)_{\pm(-1)^{nm}}~,~
{\prod}^{m}_{p=1}({\varepsilon}_pL^{\pm})\Bigr].
\label{laxreprs2-eps}
\end{eqnarray}
The flows for the functionals $\{u^{(m)}_{k,j},~v^{(m)}_{k,j}\}$,
corresponding to eqs. \p{laxreprs2}, are
\begin{eqnarray}
D^{+}_n u^{(2m)}_{k,j}&=&\sum^{n}_{p=0}
(u^{(n)}_{p,j}u^{(2m)}_{k-p+n,j-p+n} \nonumber\\
&-&(-1)^{(p+n)(k-p+n)}u^{(n)}_{p,j-k+p-n+2m}u^{(2m)}_{k-p+n,j}),
\label{equationss1}
\end{eqnarray}
\begin{eqnarray}
D^{+}_{2n} u^{(2m+1)}_{k,j}&=&\sum^{2n}_{p=0}
((-1)^{p}u^{(2n)}_{p,j}u^{(2m+1)}_{k-p+2n,j-p+2n} \nonumber\\
&-&(-1)^{p(k-p)}u^{(2n)}_{p,j-k+p-2n+2m+1}u^{(2m+1)}_{k-p+2n,j}),
\label{equationss1a}
\end{eqnarray}
\begin{eqnarray}
D^{+}_{2n+1} u^{(2m+1)}_{k,j}&=&\sum^{k}_{p=1}
((-1)^{p+1}u^{(2n+1)}_{p+2n+1,j}u^{(2m+1)}_{k-p,j-p} \nonumber\\
&+&(-1)^{p(k-p)}u^{(2n+1)}_{p+2n+1,j-k+p+2m+1}u^{(2m+1)}_{k-p,j}),
\label{equationss1b}
\end{eqnarray}
\begin{eqnarray}
D^{-}_n u^{(m)}_{k,j}&=&\sum^{n-1}_{p=0}
((-1)^{(p+n)m}v^{(n)}_{p,j}u^{(m)}_{k+p-n,j+p-n} \nonumber\\
&-&(-1)^{(p+n)(k+p-n)}v^{(n)}_{p,j-k-p+n+m}u^{(m)}_{k+p-n,j}),
\label{equationss2}
\end{eqnarray}
\begin{eqnarray}
D^{+}_n v^{(m)}_{k,j}&=&\sum^{n}_{p=0}
((-1)^{(p+n)m}u^{(n)}_{p,j}v^{(m)}_{k+p-n,j-p+n} \nonumber\\
&-&(-1)^{(p+n)(k+p-n)}u^{(n)}_{p,j+k+p-n-m}v^{(m)}_{k+p-n,j}),
\label{equationss3}
\end{eqnarray}
\begin{eqnarray}
D^{-}_n v^{(2m)}_{k,j}&=&\sum^{n-1}_{p=0}
(v^{(n)}_{p,j}v^{(2m)}_{k-p+n,j+p-n} \nonumber\\
&-&(-1)^{(p+n)(k-p+n)}v^{(n)}_{p,j+k-p+n-2m}v^{(2m)}_{k-p+n,j}),
\label{equationss4}
\end{eqnarray}
\begin{eqnarray}
D^{-}_{2n} v^{(2m+1)}_{k,j}&=&\sum^{2n-1}_{p=0}
((-1)^{p}v^{(2n)}_{p,j}v^{(2m+1)}_{k-p+2n,j+p-2n} \nonumber\\
&-&(-1)^{p(k-p)}v^{(2n)}_{p,j+k-p+2n-2m-1}v^{(2m+1)}_{k-p+2n,j}),
\label{equationss4a}
\end{eqnarray}
\begin{eqnarray}
D^{-}_{2n+1} v^{(2m+1)}_{k,j}&=&\sum^{k}_{p=0}
((-1)^{p+1}v^{(2n+1)}_{p+2n+1,j}v^{(2m+1)}_{k-p,j+p} \nonumber\\
&+&(-1)^{p(k-p)}v^{(2n+1)}_{p+2n+1,j+k-p-2m-1}v^{(2m+1)}_{k-p,j}),
\label{equationss4b}
\end{eqnarray}
where all fields $\{u^{(m)}_{k,j},~v^{(m)}_{k,j}\}$
in the right-hand side should be put equal to zero at $k < 0$.
When deriving eqs. \p{equationss1b} and \p{equationss4b}
we have used the following identity:
$2(L^{\pm})^{2(n+m+1)}_{*} = ((L^{\pm})^{2n+1}_{*})^{*}(L^{\pm})^{2m+1}_{*}
+((L^{\pm})^{2m+1}_{*})^{*}(L^{\pm})^{2n+1}_{*}$ which can easily be
verified using the definitions \p{rule2}.

\subsection{Bosonic symmetries of $N=(2|2)$ 2DTL equation}

The $N=(2|2)$ supersymmetric 2DTL equation belongs to the system of
equations (\ref{equationss1}--\ref{equationss4b}). In order to see that,
let us consider eqs. \p{equationss2} at $\{n=m=k=1\}$
\begin{eqnarray}
D^{-}_1 u_{1,j} = - v_{0,j} - v_{0,j+1}
\label{todabs1}
\end{eqnarray}
and eqs. \p{equationss3} at $\{n=m=1, k=0\}$
\begin{eqnarray}
D^{+}_1 v_{0,j}= v_{0,j}(u_{1,j} - u_{1,j-1}).
\label{todabs2}
\end{eqnarray}
Then, eliminating the field $u_{1,j}$ from eqs.
(\ref{todabs1}--\ref{todabs2}) we obtain
\begin{eqnarray}
D^{+}_1D^{-}_1 \ln v_{0,j}= v_{0,j+1} - v_{0,j-1}.
\label{todabs}
\end{eqnarray}
Equation \p{todabs} reproduces the $N=(1|1)$ superfield form of the
$N=(2|2)$ superconformal 2DTL equation (see, e.g. refs. \cite{eh,lds} and
references therein) which is the supersymmetrization of the system
of two decoupled 2DTL equations \p{todab}. Indeed, it terms of the
superfield components
\begin{eqnarray}
f_{j} \equiv v_{0,j}|,
\quad {\gamma}^{\pm}_{j} \equiv ({\cal D}^{\pm}_1\ln v_{0,j})|,
\label{suptod1}
\end{eqnarray}
where $f_j$ (${\gamma}_j,{\overline {\gamma}}_j$) are bosonic (fermionic)
fields and $|$ means the $t^{+}_1\rightarrow 0$ limit,
eq. \p{todabs} becomes
\begin{eqnarray}
{\partial}^{+}_1{\partial}^{-}_1 \ln f_j &=&
-f_{j+1}f_{j+2}+f_{j}(f_{j+1}+f_{j-1})-f_{j-1}f_{j-2} \nonumber\\
&-&f_{j+1}{\gamma}^{+}_{j+1}{\gamma}^{-}_{j+1}
+f_{j-1}{\gamma}^{+}_{j-1}{\gamma}^{-}_{j-1}, \nonumber\\
\mp {\partial}_{\mp} {\gamma}^{\pm}_{j}&=&f_{j+1}{\gamma}^{\mp}_{j+1}-
f_{j-1}{\gamma}^{\mp}_{j-1}.
\label{trlaxeqt2}
\end{eqnarray}
Then, denoting $v_j:=f_{j-1}f_j$ in the bosonic limit when all fermionic
fields are set to zero, we finally obtain the equation
\begin{eqnarray}
\partial^{-}_1\partial^{+}_1\ln v_{j}= -v_{j+2} +2 v_{j} - v_{j-2}
\label{todabsb}
\end{eqnarray}
which obviously splits into the system of two decoupled 2DTL
equations \p{todab} for the functions at even and odd lattice points,
i.e. $v_{2j}$ and $v_{2j+1}$.

Now, we would like to discuss how bosonic symmetries of the $N=(2|2)$ 2DTL
equation \p{todabs} originate from the system
(\ref{equationss1}--\ref{equationss4b}). It appears that the approach
developed in the previous section for the case of symmetries of the
bosonic 2DTL equation \p{todab} can straightforwardly be extended to the
present case.

First, let us derive the flows $D^{+}_n v_{0,j}$ and $D^{-}_n v_{0,j}$ of
the STL hierarchy considering eqs. \p{equationss3} at $\{m=1, k=0\}$
and eqs. (\ref{equationss4a}--\ref{equationss4b})
at $\{m=k=0\}$,
\begin{eqnarray}
D^{+}_n v_{0,j} = + v_{0,j}(u^{(n)}_{n,j}-u^{(n)}_{n,j-1})
\label{flowss+}
\end{eqnarray}
and
\begin{eqnarray}
D^{-}_n v_{0,j} =  -v_{0,j}(v^{(n)}_{n,j}-v^{(n)}_{n,j-1}),
\label{flowss-}
\end{eqnarray}
respectively. When calculating eqs. \p{flowss-}, we have used the relation
\begin{eqnarray}
\sum^{2n}_{p=0}((-1)^{p}v^{(2n)}_{p,j}v_{2n-p,j+p-2n} -
(-1)^{p^2}v^{(2n)}_{p,j+2n-p-1}v_{2n-p,j})=0
\label{identitys}
\end{eqnarray}
resulting from the identitity
$((L^-)^{2n}_{*})^{*}L^{-}-L^{-}(L^-)^{2n}_{*}=0$ following from the
definitions \p{rule2}.

         From the algebra \p{algebras1} one can easily observe that only
bosonic flows
$D^{\pm}_{2n}$ commute simultaneously with both the fermionic derivatives
$D^{+}_1$ and $D^{-}_1$ entering into the $N=(2|2)$ 2DTL equation \p{todabs},
\begin{eqnarray}
[D^{+}_{1}~,~D^{\pm}_{2n}]=[D^{-}_{1}~,~D^{\pm}_{2n}]=0,
\label{algebras2}
\end{eqnarray}
while the fermionic flows $D^{\pm}_{2n+1}$ do not satisfy this property.
Due to this reason, the bosonic flows $D^{\pm}_{2n}$
(\ref{flowss+}--\ref{flowss-}) form symmetries of the $N=(1|1)$ 2DTL
equation \p{todabs} if one can possibly express the functionals
$u^{(2n)}_{2n,j}$ and $v^{(2n)}_{2n,j}$ in the right-hand side of eqs.
\p{flowss+} and \p{flowss-} in terms of the field $v_{0,j}$ alone,
while the fermionic flows $D^{\pm}_{2n+1}$ do not.

With the aim to express $u^{(n)}_{n,j}$ and $v^{(n)}_{n,j}$ in terms of
$v_{0,j}$, let us consider eqs. \p{equationss2} and \p{equationss3} at $n=1$
\begin{eqnarray}
D^{-}_1 u^{(m)}_{k,j}=
(-1)^{m}v_{0,j}u^{(m)}_{k-1,j-1} +
(-1)^{k}v_{0,j-k+m+1}u^{(m)}_{k-1,j},
\label{recs1}
\end{eqnarray}
\begin{eqnarray}
D^{+}_1 v^{(m)}_{k,j}-(-1)^{k} v^{(m)}_{k,j} (u_{1,j} - u_{1,j+k-m})=
(-1)^{m}v^{(m)}_{k-1,j+1}+(-1)^{k}v^{(m)}_{k-1,j}. ~~~~
\label{recs2}
\end{eqnarray}
Substituting
\begin{eqnarray}
u_{1,j} - u_{1,j+k-m}=
+D^{+}_1 \ln \prod^{m-k}_{n=1}v_{0,j+k-m+n}, \quad m>k
\label{todab2froms1}
\end{eqnarray}
derived from eq. \p{todabs2} into eq. \p{recs2} and introducing the new
basis $v^{(m)}_{k,j}$ $\Rightarrow$ ${\widetilde v}^{(m)}_{k,j}$,
according to the formulae
\begin{eqnarray}
v^{(m)}_{m,j}={\widetilde v}^{(m)}_{m,j}, \quad
v^{(m)}_{k,j}={\widetilde v}^{(m)}_{k,j}\prod^{m-k}_{n=1}v_{0,j+k-m+n}
\quad m>k,
\label{basiss1}
\end{eqnarray}
eq. \p{recs2} becomes
\begin{eqnarray}
D^{+}_1 {\widetilde v}^{(m)}_{k,j}=
(-1)^{m}v_{0,j+1}{\widetilde v}^{(m)}_{k-1,j+1} +
(-1)^{k}v_{0,j+k-m}{\widetilde v}^{(m)}_{k-1,j}, \quad m\geq k
\label{rec2basiss1}
\end{eqnarray}
and has the form similar to eq. \p{recs1}.

The equations \p{recs1} and \p{rec2basiss1} derived represent recurrent
relations which being iterated with the starting values $u^{(n)}_{0,j}=1$
\p{imp2} and ${\widetilde v}^{(n)}_{0,j}=1$ allow one to express the
functionals $u^{(n)}_{n,j}$ and $v^{(n)}_{n,j} \equiv {\widetilde
v}^{(n)}_{n,j}$ in terms of $v_{0,j}$ after the $n$-th step of the iteration
procedure.  The latter yield the bosonic symmetries $D^{+}_{2n} v_{0,j}$ and
$D^{-}_{2n} v_{0,j}$ to the $N=(1|1)$ 2DTL equation \p{todabs} via eqs.
\p{flowss+} and \p{flowss-}.

Let us remark that the $N=(2|2)$ 2DTL equation \p{todabs} possesses the
following involution:
\begin{eqnarray}
(D^{\pm}_1)^{\star}=D^{\mp}_1, \quad
v_{0,j}^{\star}= v_{0,i-j}
\label{autobs1}
\end{eqnarray}
which relates the flows \p{flowss+}, \p{recs1} with the flows
\p{flowss-}, (\ref{basiss1}--\ref{rec2basiss1}),
\begin{eqnarray}
(D^{\pm}_n)^{\star}=D^{\mp}_n, \quad
(u^{(m)}_{k,j})^{\star}= {\widetilde v}^{(m)}_{k,i-j-1}, \quad
({\widetilde v}^{(m)}_{k,j})^{\star}=u^{(m)}_{k,i-j-1},
\label{autobs2}
\end{eqnarray}
where $i\in {\bZ}$ is a fixed number. Besides the involution
(\ref{autobs1}--\ref{autobs2}), there exists also another involution
\begin{eqnarray}
(D^{\pm}_n)^{\bullet}=D^{\pm}_n, \quad
(v_{0,j})^{\bullet}= -v_{0,i-j}.
\label{autobs3}
\end{eqnarray}
Applying the latter to \p{flowss-} and \p{rec2basiss1}
and introducing the notation
\begin{eqnarray}
({\widetilde v}^{(m)}_{k,j})^{\bullet}:= u^{(m)-}_{k,i-j-1}, \quad
u^{(m)}_{k,j}:= u^{(m)+}_{k,j},
\label{autobs4}
\end{eqnarray}
the flows (\ref{flowss+}--\ref{flowss-}), \p{recs1} and \p{rec2basiss1}
finally become
\begin{eqnarray}
D^{\pm}_n v_{0,j}&=&v_{0,j}(u^{(n)\pm}_{n,j}-u^{(n)\pm}_{n,j-1}),
\quad u^{(n)\pm}_{0,j}=1, \nonumber\\
\pm D^{\mp}_1 u^{(n)\pm}_{k,j}&=&(-1)^{n}v_{0,j}u^{(n)\pm}_{k-1,j-1} -
(-1)^{k}v_{0,j-k+n+1}u^{(n)\pm}_{k-1,j}.
\label{flowss+-}
\end{eqnarray}
The bosonic flows $D^{\pm}_{2n}$, resulting from eqs. \p{flowss+-},
\begin{eqnarray}
D^{\pm}_{2n} v_{0,j}&=&v_{0,j}(u^{(2n)\pm}_{2n,j}-u^{(2n)\pm}_{2n,j-1}),
\quad u^{(2n)\pm}_{0,j}=1, \nonumber\\ \pm D^{\mp}_1
u^{(2n)\pm}_{k,j}&=&v_{0,j}u^{(2n)\pm}_{k-1,j-1} -
(-1)^{k}v_{0,j-k+2n+1}u^{(2n)\pm}_{k-1,j},
\label{flowss+-bos}
\end{eqnarray}
reproduce the bosonic solutions to the symmetry equation corresponding to the
$N=(1|1)$ 2DTL equation \p{todabs} derived in \cite{ls1} by a rather
heuristic construction, while the algebra of the bosonic symmetries
$D^{\pm}_{2n}$ \p{flowss+-bos}
\begin{eqnarray}
[D^{\pm}_{2n}~,~D^{\pm}_{2l}]=[D^{+}_{2n}~,~D^{-}_{2l}]=0
\label{algebras3}
\end{eqnarray}
resulting from eqs. \p{algebras1} was not proved there.

\subsection{Fermionic symmetries of $N=(2|2)$ 2DTL equation}

In this subsection, we discuss the origin of the fermionic symmetries,
proposed in \cite{ols1}, of the $N=(1|1)$ 2DTL equation \p{todabs} and
construct their algebra.

For completeness, we would like to start with the derivation of a close
set of equations for the functionals $u^{(2n)}_{k,j}$ aiming to reproduce
the solutions corresponding to fermionic symmetries first observed in
\cite{ols1}.

With this goal in mind, let us consider eqs. \p{equationss1} at $n=1$,
\begin{eqnarray}
D^{+}_1 u^{(2n)}_{k,j} +
(-1)^{k}u^{(2n)}_{k,j}(u_{1,j-k+2n}-u_{1,j})
=u^{(2n)}_{k+1,j+1} +(-1)^{k}u^{(2n)}_{k+1,j}.
\label{equationss1new1}
\end{eqnarray}
Then, using the recursive substitution \p{recs1}, we express the functionals
$u^{(2n)}_{k+1,j}$ in the right-hand side of eqs. \p{equationss1new1}
in terms of the functionals $u^{(2n)}_{k,j}$; particularly, we also use the
relation
\begin{eqnarray}
u_{1,j-k+2n} - u_{1,j}=
(D^{-}_1)^{-1}(v_{0,j}+v_{0,j+1}-v_{0,j-k+2n}-v_{0,j-k+2n+1}),
\label{equationss1new2}
\end{eqnarray}
and as a result, we elaborate the following close equations for the
functionals $u^{(2n)}_{k,j}$ at different lattice points ($j-1$, $j$ and
$j+1$), but with the same subscript $k$
\begin{eqnarray}
&&(-1)^{k}D^{+}_1 u^{(2n)}_{k,j} + u^{(2n)}_{k,j}
(D^{-}_1)^{-1}(v_{0,j}-v_{0,j-k+2n+1}\nonumber\\
&& \quad
\quad \quad \quad \quad \quad \quad \quad \quad \quad \quad \quad \quad
+v_{0,j+1}-v_{0,j-k+2n})\nonumber\\
&&=(D^{-}_1)^{-1}(v_{0,j}u^{(2n)}_{k,j-1}
-v_{0,j-k+2n+1}u^{(2n)}_{k,j+1}\nonumber\\
&& \quad \quad \quad
+(-1)^{k}(v_{0,j+1}-v_{0,j-k+2n})u^{(2n)}_{k,j})
\label{equationss1new3}
\end{eqnarray}
which reproduce the corresponding equations derived by a
heuristic construction in \cite{ls1,ols1}. According to \cite{ls1,ols1},
equations \p{equationss1new3} can be treated as the result of
the application of the recursive chain of substitutions \p{recs1}
to the symmetry equation corresponding to the symmetries $D^{+}_{2n}$
\p{flowss+-bos} of the $N=(2|2)$ 2DTL equation \p{todabs}. In other words,
equations \p{equationss1new3} represent the consistency conditions
for the algebra \p{algebras2} realized on the shell of the $N=(2|2)$ 2DTL
equation \p{todabs}. Due to this reason, we can forget for a moment about
their hierarchy origin and discuss their solutions which will be relevant for
further consideration.

At $k=0$, equation \p{equationss1new3} possesses a very simple, constant
solution $u^{(2n)}_{0,j}=1$ \cite{ls1} which reproduces the condition
\p{imp2} for the hierarchy we started with. As it has already been
explained in the previous subsection, this solution generates a very
non-trivial solution for the functional $u^{(2n)}_{2n,j}$ via eqs.
\p{recs1} as well as the bosonic symmetry $D^{+}_{2n} v_{0,j}$ to the
$N=(1|1)$ 2DTL equation \p{todabs} via eq. \p{flowss+}.

It turns out that eq. \p{equationss1new3} possesses also a fermionic,
lattice-dependent solution at $k=-1$, namely \cite{ols1}
\begin{eqnarray}
u^{(2n)}_{-1,j}= (-1)^{j+1}{\epsilon},
\label{recsolb}
\end{eqnarray}
where ${\epsilon}$ is a dimensionless fermionic constant.
It remains to show how fermionic symmetries are being activated.
With this goal in mind, let us represent the bosonic time derivative
$D^{+}_{2n}$ corresponding to the solution \p{recsolb} and the functional
$u^{(2n)}_{k,j}$ which enter eqs. \p{flowss+}, \p{recs1}, \p{recsolb} 
and \p{algebras2} in the following form:
\begin{eqnarray}
D^{+}_{2n} := {\epsilon} {\cal D}^{+}_{2n+1}, \quad
u^{(2n)}_{k,j}:={\epsilon}{\cal U}^{(2n+1)+}_{k+1,j}
\label{evolder}
\end{eqnarray}
defining a new fermionic evolution derivative ${\cal D}^{+}_{2n+1}$ and
the functionals ${\cal U}^{(2n+1)+}_{k,j}$. Then, the fermionic constant
${\epsilon}$ enters linearly into both the sides of eqs. \p{flowss+},
\p{recs1}, \p{recsolb} and \p{algebras2} which now become
\begin{eqnarray}
{\cal D}^{\pm}_{2n+1} v_{0,j} &=& v_{0,j}({\cal U}^{(2n+1)\pm}_{2n+1,j}-
{\cal U}^{(2n+1)\pm}_{2n+1,j-1}), \quad
{\cal U}^{(2n+1)\pm}_{0,j}=(-1)^{j+1}, \nonumber\\
\pm D^{\mp}_1 {\cal U}^{(2n+1)\pm}_{k,j} &=&
-v_{0,j}{\cal U}^{(2n+1)\pm}_{k-1,j-1} +
(-1)^{k}v_{0,j-k+2n+2}{\cal U}^{(2n+1)\pm}_{k-1,j},
\label{recs1new1}
\end{eqnarray}
\begin{eqnarray}
\{D^{+}_1~,~{\cal D}^{\pm}_{2n+1}\}=
\{D^{-}_1~,~{\cal D}^{\pm}_{2n+1}\}=0.
\label{algebra2new}
\end{eqnarray}
When deriving eqs. (\ref{recs1new1}--\ref{algebra2new}) we have substituted
eqs. \p{evolder} into eqs. \p{flowss+}, \p{recs1}, \p{recsolb} and
\p{algebras2}, and additionally used the involution \p{autobs3} and
notation \p{autobs4}. Therefore, the flows ${\cal D}^{\pm}_{2m+1}$ do not
actually depend on ${\epsilon}$, so ${\epsilon}$ is an artificial parameter
which need not be introduced at all. The most important fact however is that
${\cal D}^{\pm}_{2m+1}$ anticommute with the fermionic derivatives
$D^{\pm}_1$ \p{algebra2new} entering into the $N=(2|2)$ 2DTL equation
\p{todabs} by construction, and due to this reason, they form symmetries
of the $N=(2|2)$ 2DTL equation \p{todabs}.

Although the existence of the symmetries ${\cal D}^{\pm}_{2n+1}$
\p{recs1new1} was established in \cite{ols1}, their algebra was only
conjectured by extending the algebra of a few first bosonic and fermionic
flows explicitly derived there. Now, we are ready to rigorously establish
the algebra of all the bosonic ${\cal D}^{\pm}_{2n}$ \p{flowss+-bos} and
fermionic ${\cal D}^{\pm}_{2n+1}$ \p{recs1new1} symmetries in the framework
of the developed here approach.

Our strategy comprises a few steps.

First, let us calculate the fermionic symmetry ${\cal D}^{\pm}_{1}v_{0,j}$
\p{recs1new1} and its algebra expressing the symmetry in terms of the
fermionic flow $D^{\pm}_{1}v_{0,j}$ \p{flowss+-} and using the algebra
\p{algebras1}. They are
\begin{eqnarray}
{\cal D}^{\pm}_{1}v_{0,j} \equiv (-1)^{j+1}D^{\pm}_{1}v_{0,j}
\label{recs1new3}
\end{eqnarray}
and
\begin{eqnarray}
\{{\cal D}^{\pm}_{1}~,~{\cal D}^{\pm}_{1}\}v_{0,j}=
-\{D^{\pm}_{1}~,~D^{\pm}_{1}\}v_{0,j}\equiv -2D^{\pm}_{2}v_{0,j},
\label{recs1new4}
\end{eqnarray}
respectively.

Second, we use the derived relation \p{recs1new3} in order to replace
$D^{\pm}_{1}$ by ${\cal D}^{\pm}_{1}$ in the expressions both for the bosonic
\p{flowss+-bos} and fermionic \p{recs1new1} symmetries, then transform them
to the new basis
\begin{eqnarray}
&&{\widehat u}^{(2n+1)\pm}_{k,j}:=
c_k (-1)^{(k+1)(j+1)}{\cal U}^{(2n+1)\pm}_{k,j}, \quad
{\widehat u}^{(2n)\pm}_{k,j}:= c_k (-1)^{kj}u^{(2n)\pm}_{k,j}, \nonumber\\
&&{\widehat {\cal D}}^{\pm}_{2n+1}:= c_{2n+1}{\cal D}^{\pm}_{2n+1}, \quad
{\widehat {\cal D}}^{\pm}_{2n}:= c_{2n}D^{\pm}_{2n},
\quad c_{2n}= c_{2n+1}\equiv (-1)^{n}
\label{recs1new5}
\end{eqnarray}
which is defined by a single requirement that the form of the symmetries
in this basis is as close as possible to the form of the flows
$D^{\pm}_{n}$ \p{flowss+-} of the STL hierarchy whose algebra
\p{algebras1} is known. In the new basis \p{recs1new5}, the symmetries
\p{flowss+-bos} and \p{recs1new1} as well the algebra \p{recs1new4} become
\begin{eqnarray}
&&{\widehat {\cal D}}^{\pm}_{n} v_{0,j} = v_{0,j}
({\widehat u}^{(n)\pm}_{n,j}-{\widehat u}^{(n)\pm}_{n,j-1}),
\quad {\widehat u}^{(n)\pm}_{0,j}=1, \nonumber\\
&&\pm{\widehat {\cal D}}^{\mp}_1 {\widehat u}^{(n)\pm}_{k,j}=
(-1)^{n}v_{0,j}{\widehat u}^{(n)\pm}_{k-1,j-1} +
(-1)^{k}v_{0,j-k+n+1}{\widehat u}^{(n)\pm}_{k-1,j}
\label{recs1new6}
\end{eqnarray}
and
\begin{eqnarray}
\{{\widehat {\cal D}}^{\mp}_{1}~,~
{\widehat {\cal D}}^{\mp}_{1}\}=2{\widehat {\cal D}}^{\mp}_{2},
\label{algebras1new}
\end{eqnarray}
respectively. When deriving the second line of eqs. \p{recs1new6}, we
have first acted by the fermionic derivative $D^{\mp}_{1}$ on both sides of
the second line of eqs. \p{flowss+-bos} and \p{recs1new1} and then used the
latter once more together with eqs. (\ref{recs1new3}--\ref{recs1new5}) and
\p{algebras1new}. A simple comparison allows one to immediately observe
that (\ref{recs1new6}) and (\ref{algebras1new}) coincide with the
expressions for the flows $D^{\pm}_{n}$ \p{flowss+-} and the algebra of the
derivatives $D^{\mp}_{1}$ \p{recs1new4}, respectively, where, however,
the evolution derivatives $D^{\pm}_{n}$ are replaced by
${\widehat {\cal D}}^{\pm}_{n}$. The obvious,
important consequence from this observation
is that the algebra of the evolution derivatives
${\widehat {\cal D}}^{\pm}_{n}$ has also to reproduce the algebra of the
evolution derivatives $D^{\pm}_{n}$ \p{algebras1}. Thus, we are led to the
following formulae for both this algebra and the algebras \p{recs1new4},
\p{algebras2} as well as \p{algebra2new} transformed to the basis
\p{recs1new5}
\begin{eqnarray}
[{\widehat {\cal D}}^{+}_{n}~,~
{\widehat {\cal D}}^{-}_{l}\}=
[{\widehat {\cal D}}^{\pm}_{n}~,~{\widehat {\cal D}}^{\pm}_{2l}]=0,\quad
\{{\widehat {\cal D}}^{\pm}_{2n+1}~,~
{\widehat {\cal D}}^{\pm}_{2l+1}\}=2{\widehat {\cal D}}^{\pm}_{2(n+l+1)}
\label{algebras1new1}
\end{eqnarray}
and
\begin{eqnarray}
&&\{D^{+}_{1}~,~D^{-}_{1}\}=0, \quad
\{D^{\pm}_{1}~,~D^{\pm}_{1}\}= -2{\widehat {\cal D}}^{\pm}_{2}, \nonumber\\
&& [D^{+}_1~,~{\widehat {\cal D}}^{\pm}_{n}\}=
[D^{-}_1~,~{\widehat {\cal D}}^{\pm}_{n}\}=0.
\label{algebras1neww1}
\end{eqnarray}

By construction, the algebra \p{algebras1new1} forms a symmetry algebra
of the $N=(2|2)$ 2DTL equation \p{todabs}. However, one can easily understand
that the fermionic symmetries ${\cal D}^{\pm}_{2n+1}$ \p{recs1new1} are
also symmetries of the bosonic flows $D^{\pm}_{2n}$ \p{flowss+-bos} of the
STL hierarchy because of the following commutation relations:
\begin{eqnarray}
[{\cal D}^{+}_{2n+1}~,~D^{\pm}_{2l}]=
[{\cal D}^{-}_{2n+1}~,~D^{\pm}_{2l}]=0
\label{algebras1new2}
\end{eqnarray}
resulting from the algebra \p{algebras1new1} and the relations \p{recs1new5}.

Let us also point out that bosonic and fermionic symmetries of the
one-dimensional reduction of the $N=(2|2)$ 2DTL hierarchy --- the $N=4$
supersymmetric Toda chain hierarchy --- were analyzed in detail in
\cite{dgs,ds}. 

The existence of the fermionic symmetries ${\cal D}^{\pm}_{2n+1}$
\p{recs1new1} means that the Lax pair equations \p{laxreprs1}, we started
with in this section, are not complete because they do not contain
fermionic flows which would correspond to these symmetries. Therefore, the
new problem arises: it would be interesting to construct both additional
evolution equations for the Lax operators $L^{\pm}$ \p{laxs1} generated by
the fermionic symmetries ${\cal D}^{\pm}_{2n+1}$ \p{recs1new1} and
commutation relations between the latter and the fermionic flows
$D^{\pm}_{2n+1}$ \p{flowss+-} of the STL hierarchy. The detailed analysis of
this rather nontrivial problem is beyond the scope of the present paper and
will be considered elsewhere. Let us only mention that a similar task
has partly been discussed in \cite{t} in a slightly different context.

To close this section, let us briefly discuss one of the consequences of the
results derived in this subsection which is important in the context of the
problem of constructing an $N=(2|2)$ superfield formulation of the bosonic
flows $D^{\pm}_{2n}$ \p{flowss+-bos} of the STL hierarchy. Quite recently
this problem was considered in \cite{ols2} basing on the conjecture partly
proved there (for more details, see ref. \cite{ols2}). In terms of the
objects introduced in the present paper this conjecture can be reformulated
as a conjecture about the validity of the following constraints:
\begin{eqnarray}
\Bigl({\cal D}^{\mp}_{1} + D^{\mp}_{1}\Bigr)D^{\pm}_{2l}~v_{0,2j}=0, \quad
\Bigl({\cal D}^{\mp}_{1} - D^{\mp}_{1}\Bigr)D^{\pm}_{2l}~v_{0,2j+1}=0,
\label{conjecture1a}
\end{eqnarray}
\begin{eqnarray}
\Bigl({\cal D}^{\pm}_{1} + D^{\pm}_{1}\Bigr)D^{\pm}_{2l}~v_{0,2j}=0, \quad
\Bigl({\cal D}^{\pm}_{1} - D^{\pm}_{1}\Bigr)D^{\pm}_{2l}~v_{0,2j+1}=0.
\label{conjecture1b}
\end{eqnarray}
The proof that the constraints \p{conjecture1a} are in fact satisfied is
given in \cite{ols2}. As concerns the remaining constraints
\p{conjecture1b}, only evidence in their favour was presented there by
confirming them (and \p{conjecture1a}) explicitly for the first three bosonic
flows $D^{\pm}_{2n}$ $(n=1,2,3)$ from the set \p{flowss+-bos}. Here, we are
ready to prove this conjecture. Thus, using the relations \p{recs1new3}
represented in the equivalent form
\begin{eqnarray}
\Bigl({\cal D}^{\mp}_{1} + D^{\mp}_{1}\Bigr)v_{0,2j}=0, \quad
\Bigl({\cal D}^{\mp}_{1} - D^{\mp}_{1}\Bigl)v_{0,2j+1}=0,
\label{conjecture2c}
\end{eqnarray}
the constraints (\ref{conjecture1a}--\ref{conjecture1b}) can identically be
rewritten in the following form more convenient for a further analysis:
\begin{eqnarray}
[{\cal D}^{\mp}_{1} + D^{\mp}_{1}~,~D^{\pm}_{2l}]~v_{0,2j}=0, \quad
[{\cal D}^{\mp}_{1} - D^{\mp}_{1}~,~D^{\pm}_{2l}]~v_{0,2j+1}=0,
\label{conjecture2a}
\end{eqnarray}
\begin{eqnarray}
[{\cal D}^{\pm}_{1} + D^{\pm}_{1}~,~D^{\pm}_{2l}]~v_{0,2j}=0, \quad
[{\cal D}^{\pm}_{1} - D^{\pm}_{1}~,~D^{\pm}_{2l}]~v_{0,2j+1}=0.
\label{conjecture2b}
\end{eqnarray}
It is a simple exercise now to verify that the correctness of the
conjecture in the form of equations (\ref{conjecture2a}--\ref{conjecture2b})
is a direct consequence of the algebras \p{algebras2} and \p{algebras1new2}.

\section{N=(0$|$2) supersymmetric 2DTL hierarchy}

In this section we propose the new, $N=(0|2)$ supersymmetric 2DTL hierarchy
which includes the $N=(0|2)$ superconformal 2DTL equation derived in
\cite{lds} and construct both bosonic and fermionic symmetries of the
latter.

\subsection{Lax pair representation and flows}

Let us start with the following set of the consistent Lax pair
equations:
\begin{eqnarray}
D^{+}_n L^{+}&=&
(-1)^{n}(((L^{+})^{n}_{*})_{+})^{*} L^{+}
-(L^{+})^{*(n)}((L^{+})^{n}_{*})_{+} \nonumber\\
&+&(1-(-1)^n)(L^{+})^{n+1}_{*}, \nonumber\\
D^{+}_n L^{-}&=&
((L^{+})^{n}_{*})_{+} L^{-}-
(L^{-})^{*(n)}((L^{+})^{n}_{*})_{+}, \nonumber\\
D^{-}_{2n} L^{+}&=&
(((L^{-})^{n})_{-})^{*} L^{+}
-(L^{+})((L^{-})^{n})_{-}, \nonumber\\
D^{-}_{2n} L^{-}&=&[((L^{-})^{n})_{-}~,~L^{-}], \quad n \in {\bN},
\label{laxreprsf1}
\end{eqnarray}
\begin{eqnarray}
L^{+}=\sum^{\infty}_{k=0} u_{k,j}e^{(1-k){\partial}}, \quad
L^{-}=\sum^{\infty}_{k=0} v_{k,j}e^{(k-2){\partial}},
\label{laxsf1}
\end{eqnarray}
\begin{eqnarray}
u_{0,j}\equiv 1, \quad v_{0,2j+1}\equiv 0, \quad v_{0,2j}\neq 0
\label{bound3}
\end{eqnarray}
generating the non-abelian algebra of the flows
\begin{eqnarray}
[D^{+}_{n}~,~D^{\pm}_{2l}]=[D^{-}_{2n}~,~D^{-}_{2l}]=0, \quad
\{D^{+}_{2n+1}~,~D^{+}_{2l+1}\}=2D^{+}_{2(n+l+1)}
\label{algebrasf1}
\end{eqnarray}
which may be realized in the superspace  $\{t^{+}_n, t^{-}_{2n}\}$
\begin{eqnarray}
D^{\pm}_{2n} ={\partial}^{\pm}_{2n}, \quad
D^{+}_{2n+1} ={\partial}^{+}_{2n+1}+
\sum^{\infty}_{l=1}t^{+}_{2l-1}{\partial}^{+}_{2(k+l)},
\label{covderf}
\end{eqnarray}
where $D^{\pm}_{2n}$ and $t^{\pm}_{2n}$ ($D^{+}_{2n+1}$ and
$t^{+}_{2n+1}$) are bosonic (fermionic) evolution derivatives
and times, respectively;
$u_{2k,j}(\{t^{+}_n,t^{-}_{2n}\})$ and
$v_{2k,j}(\{t^{+}_n,t^{-}_{2n}\})$ \\
( $u_{2k+1,j}(\{t^{+}_n,t^{-}_{2n}\})$ and
$v_{2k+1,j}(\{t^{+}_n,t^{-}_{2n}\})$ )
are bosonic (fermionic) lattice fields.

In what follows we will show that the $N=(0|2)$ 2DTL equation \cite{lds}
belongs to the set of equations \p{laxreprsf1} and due to this reason we call
it the $N=(0|2)$ supersymmetric 2DTL hierarchy.

Let us introduce the following useful notation:
\begin{eqnarray}
(L^{+})^{m}_{*}:=\sum^{\infty}_{k=0} u^{(m)}_{k,j}e^{(m-k){\partial}}, \quad
(L^{-})^{m}:=\sum^{\infty}_{k=0} v^{(m)}_{k,j}e^{(k-2m){\partial}}
\label{laxsf2}
\end{eqnarray}
which will be used in this section. Here,
$\{u^{(m)}_{2k,j},~v^{(m)}_{2k,j}\}$  and
$\{u^{(m)}_{2k+1,j},~v^{(m)}_{2k+1,j}\}$ ($u^{(1)}_{k,j}\equiv
u_{k,j},~v^{(1)}_{k,j}\equiv v_{k,j}$) are bosonic and fermionic functionals
of the original fields $\{u_{k,j},~v_{k,j}\}$ whose explicit form is not
important for the further consideration but the explicit form of
the following functionals:
\begin{eqnarray}
u^{(m)}_{0,j}=1, \quad v^{(m)}_{0,2j+1}= 0, \quad v^{(m)}_{0,2j}\neq 0
\label{bound4}
\end{eqnarray}
which can easily be found using eqs. \p{bound3}.

One important remark is in order: the Lax pair representation
(\ref{laxreprsf1}--\ref{laxsf1}) supplied by the constraints \p{bound3}
cannot be obtained by reducting the Lax pair representation
(\ref{laxreprs1}--\ref{bound2}) of the $N=(2|2)$ 2DTL hierarchy.
Indeed, if it would be the case, then the Lax operator $L^{-}$ \p{laxsf1}
had the square root of the form
\begin{eqnarray}
(L^{-})=((L^{-})^{\frac{1}{2}})^{*}(L^{-})^{\frac{1}{2}}, \quad
(L^{-})^{\frac{1}{2}}=
\sum^{\infty}_{k=0} v^{(\frac{1}{2})}_{k,j}e^{(k-1){\partial}}
\label{sroot1}
\end{eqnarray}
which reproduces the original Lax operator \p{laxs1}
of the $N=(2|2)$ 2DTL hierarchy,
and as a consequence of eqs. \p{sroot1}, the field $v_{0,j}$
admits the following representation:
\begin{eqnarray}
v_{0,j}=v^{(\frac{1}{2})}_{0,j}v^{(\frac{1}{2})}_{0,j-1}.
\label{sroot2}
\end{eqnarray}
However, the latter is inconsistent with the conditions \p{bound3}; so
we come to the contradiction. Therefore, the conclusion is that
the $N=(2|2)$ 2DTL hierarchy cannot be reduced to the $N=(0|2)$ 2DTL
hierarchy.

The following operator equations:
\begin{eqnarray}
D^{+}_n (L^{+})^{m}_{*}&=&
(-1)^{nm}(((L^{+})^{n}_{*})_{+})^{*(m)} (L^{+})^{m}_{*} \nonumber\\
&-&((L^{+})^{m}_{*})^{*(n)}((L^{+})^{n}_{*})_{+}
+(1-(-1)^n)(L^{+})^{n+m}_{*}, \nonumber\\
D^{+}_n (L^{-})^{m}&=&((L^{+})^{n}_{*})_{+} (L^{-})^{m}-
((L^{-})^{m})^{*(n)}((L^{+})^{n}_{*})_{+},
\nonumber\\ D^{-}_{2n} (L^{+})^{m}_{*}&=&
(((L^{-})^{n})_{-})^{*(m)} (L^{+})^{m}_{*}
-((L^{+})^{m}_{*})((L^{-})^{n})_{-}, \nonumber\\
D^{-}_{2n} (L^{-})^{m}&=&[((L^{-})^{n})_{-}, (L^{-})^{m}],
\quad n,m \in {\bN}
\label{laxreprsf2}
\end{eqnarray}
are identically satisfied on the shell of the original equations
\p{laxreprsf1}, and the corresponding flows for the
functionals $\{u^{(m)}_{k,j},~v^{(m)}_{k,j}\}$ are
\begin{eqnarray}
D^{+}_n u^{(2m)}_{k,j}&=&\sum^{n}_{p=0}
(u^{(n)}_{p,j}u^{(2m)}_{k-p+n,j-p+n} \nonumber\\
&-&(-1)^{(p+n)(k-p+n)}u^{(n)}_{p,j-k+p-n+2m}u^{(2m)}_{k-p+n,j}),
\label{equationssf1}
\end{eqnarray}
\begin{eqnarray}
D^{+}_{2n} u^{(2m+1)}_{k,j}&=&\sum^{2n}_{p=0}
((-1)^{p}u^{(2n)}_{p,j}u^{(2m+1)}_{k-p+2n,j-p+2n} \nonumber\\
&-&(-1)^{p(k-p)}u^{(2n)}_{p,j-k+p-2n+2m+1}u^{(2m+1)}_{k-p+2n,j}),
\label{equationssf1a}
\end{eqnarray}
\begin{eqnarray}
D^{+}_{2n+1} u^{(2m+1)}_{k,j}&=&\sum^{k}_{p=1}
((-1)^{p+1}u^{(2n+1)}_{p+2n+1,j}u^{(2m+1)}_{k-p,j-p} \nonumber\\
&+&(-1)^{p(k-p)}u^{(2n+1)}_{p+2n+1,j-k+p+2m+1}u^{(2m+1)}_{k-p,j}),
\label{equationssf1b}
\end{eqnarray}
\begin{eqnarray}
D^{-}_{2n} u^{(m)}_{k,j}&=&\sum^{2n-1}_{p=0}
((-1)^{pm}v^{(n)}_{p,j}u^{(m)}_{k+p-2n,j+p-2n} \nonumber\\
&-&(-1)^{p(k+p)}v^{(n)}_{p,j-k-p+2n+m}u^{(m)}_{k+p-2n,j}),
\label{equationssf2}
\end{eqnarray}
\begin{eqnarray}
D^{+}_n v^{(m)}_{k,j}&=&\sum^{n}_{p=0}
(u^{(n)}_{p,j}v^{(m)}_{k+p-n,j-p+n} \nonumber\\
&-&(-1)^{(p+n)(k+p-n)}u^{(n)}_{p,j+k+p-n-2m}v^{(m)}_{k+p-n,j}),
\label{equationssf3}
\end{eqnarray}
\begin{eqnarray}
D^{-}_{2n} v^{(m)}_{k,j}&=&\sum^{2n-1}_{p=0}
(v^{(n)}_{p,j}v^{(m)}_{k-p+2n,j+p-2n} \nonumber\\
&-&(-1)^{p(k-p)}v^{(n)}_{p,j+k-p+2n-2m}v^{(m)}_{k-p+2n,j}),
\label{equationssf4}
\end{eqnarray}
where all fields $\{u^{(m)}_{k,j},~v^{(m)}_{k,j}\}$
in the right-hand side should be put equal to zero at $k < 0$.

\subsection{Bosonic symmetries of $N=(0|2)$ 2DTL equation}

Now, let us demonstrate how the $N=(0|2)$ 2DTL equation and its symmetries
originate from this background.

With this goal in mind, let us consider eqs. \p{equationssf3} and
\p{equationssf4} at $\{m=1, k=0\}$ and $\{m=k=1\}$,
\begin{eqnarray}
D^{+}_n v_{0,2j} = +v_{0,2j}(u^{(n)}_{n,2j}-u^{(n)}_{n,2(j-1)}),
\label{flowssf1+}
\end{eqnarray}
\begin{eqnarray}
D^{+}_n v_{1,2j}&=& - v_{0,2j}u^{(n)}_{n-1,2(j-1)}
-(-1)^{n}v_{1,2j}(u^{(n)}_{n,2j-1}-u^{(n)}_{n,2j}), \nonumber\\
D^{+}_n v_{1,2j+1}&=&+v_{0,2(j+1)}u^{(n)}_{n-1,2j+1}
-(-1)^{n}v_{1,2j+1}(u^{(n)}_{n,2j}-u^{(n)}_{n,2j+1})
\label{flowssf3+}
\end{eqnarray}
and
\begin{eqnarray}
D^{-}_{2n} v_{0,2j} = -v_{0,2j}(v^{(n)}_{2n,2j}-v^{(n)}_{2n,2(j-1)}),
\label{flowssf1-}
\end{eqnarray}
\begin{eqnarray}
D^{-}_{2n} v_{1,2j}&=&v_{0,2j}v^{(n)}_{2n+1,2(j-1)} +
v_{1,2j}(v^{(n)}_{2n,2j-1}-v^{(n)}_{2n,2j}), \nonumber\\
D^{-}_{2n} v_{1,2j+1}&=&-v_{0,2(j+1)}v^{(n)}_{2n+1,2j+1} +
v_{1,2j+1}(v^{(n)}_{2n,2j}-v^{(n)}_{2n,2j+1}), ~~~
\label{flowssf3-}
\end{eqnarray}
respectively, which involve the two fields, $v_{0,2j}$ and
$v_{1,j}$. When deriving these equations we have
used the conditions \p{bound3} and the relation
\begin{eqnarray}
\sum^{k}_{p=0}(v^{(n)}_{p,j}v_{k-p,j+p-2n} -
v_{k-p,j}v^{(n)}_{p,j+k-p-2})=0
\label{identitysf}
\end{eqnarray}
at $k=2n$ and $k=2n+1$ which is a direct consequence of the identitity
$(L^-)^nL^{-}-L^{-}(L^-)^n=0$.
Equations (\ref{flowssf1+}--\ref{flowssf3-}) can further be simplified
if one introduces the new basis $\{v_{0,2j}~,~v_{1,2j}~,~v_{1,2j-1}\}$
$\Rightarrow$ $\{g_j, F_j, {\overline F}_j\}$, according to the formulae
\begin{eqnarray}
v_{0,2j}=g_{2j}g_{2j-1}, \quad v_{1,2j} = g_{2j} F_j, \quad
v_{1,2j-1} = g_{2j-1} {\overline F}_j,
\label{basis2}
\end{eqnarray}
where $F_j, {\overline F}_j$ ($g_j$) are new fermionic (bosonic) fields.
Then eqs. (\ref{flowssf1+}--\ref{flowssf3-}) become
\begin{eqnarray}
D^{+}_n \ln g_j = u^{(n)}_{n,j}-u^{(n)}_{n,j-1},
\label{flowssf1+1}
\end{eqnarray}
\begin{eqnarray}
D^{+}_n F_j= - g_{2j-1}u^{(n)}_{n-1,2(j-1)}, \quad
D^{+}_n {\overline F}_j= + g_{2j}u^{(n)}_{n-1,2j-1}
\label{flowssf2+1}
\end{eqnarray}
and
\begin{eqnarray}
D^{-}_{2n} \ln g_j = -v^{(n)}_{2n,j}+v^{(n)}_{2n,j-1},
\label{flowssf1-1}
\end{eqnarray}
\begin{eqnarray}
D^{-}_{2n} F_j= + g_{2j-1}v^{(n)}_{2n+1,2(j-1)}, \quad
D^{-}_{2n} {\overline F}_j= - g_{2j}v^{(n)}_{2n+1,2j-1}.
\label{flowssf2-1}
\end{eqnarray}
Now, using eqs. \p{bound4} one can resolve eqs. \p{flowssf2+1} at $n=1$ and
express the field $g_j$ in terms of the fields $F_j,{\overline F}_j$,
\begin{eqnarray}
g_{2j-1}=-D^{+}_1 F_j, \quad g_{2j}=+D^{+}_1 {\overline F}_j.
\label{resol1}
\end{eqnarray}
Finally, eliminating $g_j$ \p{resol1} from eqs. \p{flowssf1+1}
and \p{flowssf1-1} we obtain the following set of equations for the
fields $F_j,{\overline F}_j$:
\begin{eqnarray}
D^{+}_n \ln D^{+}_1 {\overline F}_j = u^{(n)}_{n,2j}-u^{(n)}_{n,2j-1},
\quad D^{+}_n \ln D^{+}_1 F_j = u^{(n)}_{n,2j-1}-u^{(n)}_{n,2(j-1)} ~~~~~
\label{flows+F}
\end{eqnarray}
and
\begin{eqnarray}
D^{-}_{2n}\ln D^{+}_1 {\overline F}_j &=& -v^{(n)}_{2n,2j}+v^{(n)}_{2n,2j-1},
\nonumber\\ D^{-}_{2n}\ln D^{+}_1 F_j &=&
-v^{(n)}_{2n,2j-1}+v^{(n)}_{2n,2(j-1)}.
\label{flows-F}
\end{eqnarray}
Alternatively, substituting $g_j$ from eqs. \p{resol1} into eqs.
\p{flowssf2+1} and \p{flowssf2-1} we have
\begin{eqnarray}
D^{+}_n F_j= (D^{+}_1 F_j)u^{(n)}_{n-1,2(j-1)}, \quad
D^{+}_n {\overline F}_j= (D^{+}_1 {\overline F}_j)u^{(n)}_{n-1,2j-1}
\label{flows+Fa}
\end{eqnarray}
and
\begin{eqnarray}
D^{-}_{2n} F_j&=& -(D^{+}_1 F_j)v^{(n)}_{2n+1,2(j-1)}, \nonumber\\
D^{-}_{2n} {\overline F}_j&=&- (D^{+}_1 {\overline F}_j)v^{(n)}_{2n+1,2j-1}.
\label{flows-Fa}
\end{eqnarray}

Now, it is necessary to express the functionals $u^{(n)}_{k,j}$ and
$v^{(n)}_{k,j}$ entering into the right-hand sides of eqs.
(\ref{flows+F}--\ref{flows-F}) (or eqs. (\ref{flows+Fa}--\ref{flows-Fa}) )
in terms of the fields $\{F_j,{\overline F}_j\}$ in order to have
a closed set of equations for the latter. With this goal in mind, let
us consider eqs. \p{equationssf2} and eqs. \p{equationssf3} at $n=1$,
\begin{eqnarray}
D^{-}_{2} u^{(m)}_{k,j}&=& v_{0,j}u^{(m)}_{k-2,j-2} -
v_{0,j-k+m+2}u^{(m)}_{k-2,j} \nonumber\\
&+&(-1)^{m}v_{1,j}u^{(m)}_{k-1,j-1} +
(-1)^{k}v_{1,j-k+m+1}u^{(m)}_{k-1,j}
\label{recsf1}
\end{eqnarray}
and
\begin{eqnarray}
D^{+}_1 v^{(m)}_{k,j}- (-1)^{k}v^{(m)}_{k,j} (u_{1,j}-u_{1,j+k-2m})=
v^{(m)}_{k-1,j+1} + (-1)^{k}v^{(m)}_{k-1,j}, ~~~
\label{recsf2}
\end{eqnarray}
where $v_{0,j}$ and $v_{1,j}$
should be expressed in terms of $\{F_j,{\overline F}_j\}$
using eqs. \p{bound3}, \p{basis2} and \p{resol1},
\begin{eqnarray}
&&v_{0,2j+1}=0, \quad \quad \quad
v_{0,2j}=-(D^{+}_1F_j)D^{+}_1{\overline F}_j, \nonumber\\
&&v_{1,2j} =  F_jD^{+}_1 {\overline F}_j, \quad
v_{1,2j-1} = -(D^{+}_1 F_j) {\overline F}_j.
\label{resol2}
\end{eqnarray}
Substituting
\begin{eqnarray}
u_{1,j} - u_{1,j+k-2m}= +D^{+}_1 \ln
\prod^{2m-k}_{n=1}g_{j+k-2m+n}, \quad 2m>k,
\label{todab2froms1f}
\end{eqnarray}
obtained from eq. \p{flowssf1+1} at $n=1$, into eq. \p{recsf2} and
introducing the new basis $v^{(m)}_{k,j}$ $\Rightarrow$
${\widetilde v}^{(m)}_{k,j}$,
\begin{eqnarray}
v^{(m)}_{2m,j}={\widetilde v}^{(m)}_{2m,j}, \quad
v^{(m)}_{k,j}={\widetilde v}^{(m)}_{k,j}\prod^{2m-k}_{n=1}g_{j+k-2m+n},
\quad 2m>k
\label{basiss1f}
\end{eqnarray}
eq. \p{recsf2} becomes simpler
\begin{eqnarray}
D^{+}_1 {\widetilde v}^{(m)}_{k,j}=
g_{j+1}{\widetilde v}^{(m)}_{k-1,j+1} +
(-1)^{k}g_{j+k-2m}{\widetilde v}^{(m)}_{k-1,j}, \quad 2m\geq k,
\label{rec2basiss1f}
\end{eqnarray}
where $g_j$ is given in terms of $\{F_j,{\overline F}_j\}$
by eqs. \p{resol1}.

The equations \p{recsf1} and \p{rec2basiss1f} derived represent
recurrent relations connecting the functional $u^{(n)}_{k,j}$ and
${\widetilde v}^{(n)}_{k,j}$ with $\{u^{(n)}_{k-1,i}~,~u^{(n)}_{k-2,i}\}$
and ${\widetilde v}^{(n)}_{k-1,i}$, respectively. Being iterated with the
starting values $\{u^{(n)}_{-1,j}=0, u^{(n)}_{0,j}=1\}$ \p{bound4} and
${\widetilde v}^{(n)}_{0,j}=1$, respectively, they allow one to express the
functionals $u^{(n)}_{n,j}$ and $v^{(n)}_{2n,j} \equiv {\widetilde
v}^{(n)}_{2n,j}$ in terms of $\{F_j,{\overline F}_j\}$ after the $n$-th and
$2n$-th steps of the iteration procedure, respectively . The latter yield
the flows $D^{+}_n$ and $D^{-}_{2n}$ of the fields $F_j$ and
${\overline F}_j$ via eqs. \p{flows+F} and \p{flows-F}.

For illustration, we present explicitly the flows $D^{+}_n$ \p{flows+F}
at $n=1$, $n=2$ and $n=4$ constructed by the above-described
algorithmic procedure which allows one to pass step by step,
\begin{eqnarray}
D^{-}_{2} D^{+}_1 \ln D^{+}_1F_{j+1}&=& +F_jD^{+}_1{\overline F}_j -
F_{j+1}D^{+}_1{\overline F}_{j+1}, \nonumber\\
D^{-}_{2} D^{+}_1 \ln D^{+}_1{\overline F}_{j}&=& -(D^{+}_1F_j)
{\overline F}_j +(D^{+}_1 F_{j+1}){\overline F}_{j+1},
\label{f-toda-n=1}
\end{eqnarray}
\begin{eqnarray}
D^{+}_2 F_{j}=(D^{+}_1)^2F_{j}, \quad
D^{+}_2 {\overline F}_j=(D^{+}_1)^2{\overline F}_j,
\label{n=2}
\end{eqnarray}
\begin{eqnarray}
D^{+}_4 F_{j}&=&-(D^{+}_1)^4F_{j}-
2(D^{+}_1 F_{j})(D^{-}_{2})^{-1}(D^{+}_1)^2(F_jD^{+}_1{\overline F}_j)
\nonumber\\
&+&2((D^{+}_1)^{2} F_{j})(D^{-}_{2})^{-1}(D^{+}_1)^2(F_j{\overline F}_j),
\nonumber\\
D^{+}_4 {\overline F}_j&=&+(D^{+}_1)^4{\overline F}_j
+2(D^{+}_1 {\overline F}_j)(D^{-}_{2})^{-1}(D^{+}_1)^2((D^{+}_1F_j)
{\overline F}_j)\nonumber\\
&+& 2((D^{+}_1)^{2}
{\overline F}_j)(D^{-}_{2})^{-1}(D^{+}_1)^2(F_j{\overline F}_j).
\label{ds-eq}
\end{eqnarray}
When deriving eqs. (\ref{n=2}--\ref{ds-eq}) we have used eqs. \p{f-toda-n=1}
in order to express the fields $\{F_{j+i}~,~{\overline F}_{j+i}\}$,
appearing at different lattice points $j+i$, in terms of the fields
$\{F_{j}~,~{\overline F}_{j}\}$ at the lattice point $j$.

Equations \p{f-toda-n=1} reproduce the $N=(1|1)$ superfield form of the
$N=(0|2)$ superconformal 2DTL equation \cite{lds} which is the minimal
supersymmetrization of the 2DTL equation \p{todab}. Let us discuss
this point in more detail. Thus, in terms
of the superfield components
\begin{eqnarray}
V_j \equiv D^{+}_1 {\overline F}_j|, \quad
{\overline {\Psi}}_j \equiv  {\overline F}_j|, \quad
U_j  \equiv D^{+}_1F_j|, \quad {\Psi}_j \equiv  F_j|,
\label{com1}
\end{eqnarray}
where $U_j,V_j$ (${\Psi}_j,{\overline {\Psi}}_j$) are bosonic (fermionic)
fields and $|$ means the $t^{+}_1\rightarrow 0$ limit, eqs. \p{f-toda-n=1}
become
\begin{eqnarray}
&& {\partial}^{+}_2
\Bigr({\partial}^{-}_2\ln (U_{j}V_{j-1})-\Psi_{j}\overline \Psi_{j}+
\Psi_{j-1} \overline \Psi_{j-1}\Bigl)=0, \nonumber\\
&& {\partial}^{-}_2(\frac{1}{U_j} {\partial}^{+}_2\Psi_{j})
=V_{j-1}\Psi_{j-1}-V_{j}\Psi_{j},\nonumber\\
&&{\partial}^{-}_2(\frac{1}{V_j} {\partial}^{+}_2\overline \Psi_j)=
U_{j+1} \overline \Psi_{j+1}-U_{j}\overline \Psi_{j}, \nonumber\\
&&{\partial}^{-}_2{\partial}^{+}_2 \ln V_j =  U_{j+1}V_{j+1}-U_{j}V_{j}+
({\partial}^{+}_2\Psi_{j+1}) \overline \Psi_{j+1}
-({\partial}^{+}_2\Psi_j)\overline \Psi_j. ~~~~
\label{trlaxeqt4}
\end{eqnarray}
The first equation of system \p{trlaxeqt4} has the form of a
conservation law with respect to the coordinate $t^{+}_2$.
Resolving this equation in the form
\begin{eqnarray}
{\partial}^{-}_2\ln (U_{j}V_{j-1})-\Psi_{j}\overline \Psi_{j}+
\Psi_{j-1} \overline \Psi_{j-1}={\partial}_+\ln
({\eta}_{j-1}(t^{-}_2)/{\eta}_{j}(t^{-}_2))
\label{conserv1}
\end{eqnarray}
and rescaling the fields
\begin{eqnarray}
u_j := {\eta}_{j} U_j, \quad
v_j := {V_j \over {\eta}_{j}}, \quad
\psi_j:= {\eta}_{j} {\Psi_j}, \quad
\overline \psi_j := {\overline \Psi_j \over {\eta}_{j}}
\label{condtr1}
\end{eqnarray}
we rewrite equations \p{trlaxeqt4} in an equivalent component
\begin{eqnarray}
&& {\partial}^{-}_2\ln (u_{j}v_{j-1})=\psi_{j}\overline \psi_{j}-
\psi_{j-1} \overline \psi_{j-1}, \nonumber\\
&& {\partial}^{-}_2(\frac{1}{u_j} {\partial}^{+}_2\psi_{j})
=v_{j-1}\psi_{j-1}-v_{j}\psi_{j},\nonumber\\
&&{\partial}^{-}_2(\frac{1}{v_j} {\partial}^{+}_2\overline \psi_j)=
u_{j+1} \overline \psi_{j+1}-u_{j}\overline \psi_{j}, \nonumber\\
&&{\partial}^{-}_2{\partial}^{+}_2 \ln v_j =  u_{j+1}v_{j+1}-u_{j}v_{j}+
({\partial}^{+}_2\psi_{j+1}) \overline \psi_{j+1}
-({\partial}^{+}_2\psi_j)\overline \psi_j ~~~
\label{trlaxeq6}
\end{eqnarray}
and superfield
\begin{eqnarray}
D^{-}_{2} \ln \Bigl((D^{+}_1F_{j+1})(D^{+}_1{\overline F})\Bigr)
&=& -F_j{\overline F}_j + F_{j+1}{\overline F}_{j+1}, \nonumber\\
D^{-}_{2} D^{+}_1 \ln D^{+}_1{\overline F}_{j}&=& -(D^{+}_1F_j)
{\overline F}_j +(D^{+}_1 F_{j+1}){\overline F}_{j+1}
\label{f-toda-n=1n}
\end{eqnarray}
form where an arbitrary function ${\eta}_j(t^{-}_2)$, introduced in eq.
\p{conserv1}, completely disappears. The equations \p{trlaxeq6} reproduce the
component form of the $N=(0|2)$ 2DTL equation \cite{lds} which can be reduced
to the one--dimensional $N=2$ supersymmetric Toda chain equations
\cite{ls2} by the reduction constraint ${\partial}^{+}_2={\partial}^{-}_2$.
Let us also point out that bosonic and fermionic symmetries of this reduction
were analyzed in detail in \cite{bs}. In the bosonic limit, when all fermionic
fields are set to zero, equations \p{trlaxeq6} become
\begin{eqnarray}
{\partial}^{-}_2\ln (u_{j}v_{j-1})=0, \quad
{\partial}^{-}_2{\partial}^{+}_2 \ln v_j =  u_{j+1}v_{j+1}-u_{j}v_{j},
\label{trlaxeq6bos}
\end{eqnarray}
and the equation, resulting obviously from them, for the function
$b_j\equiv -u_j v_j$
\begin{eqnarray}
\partial^{-}_2\partial^{+}_2\ln b_{j}= -b_{j+1} +2 b_{j} - b_{j-1}
\label{todabnew}
\end{eqnarray}
reproduces the 2DTL equation \p{todab}.

As concerns eqs. \p{ds-eq}, they represent minimal
supersymmetrization of the Davey-Stewartson equation \cite{d-s} which is the
$(2+1)$-dimensional generalization of the $(1+1)$-dimensional Nonlinear
Schroedinger equation.

Let us remark that the $N=(0|2)$ 2DTL equation \p{f-toda-n=1}
as well as the equations (\ref{n=2}--\ref{ds-eq}) possess the
following involution:
\begin{eqnarray}
(F_{j})^{\star}= {\overline F}_{i-j}, \quad
({\overline F}_{i-j})^{\star}= F_{i-j},
\label{autobsf1}
\end{eqnarray}
where $i\in {\bZ}$ is an arbitrary fixed number.

         From the algebra \p{algebrasf1} we learn that only bosonic flows
$D^{\pm}_{2n}$ of the $N=(0|2)$ 2DTL hierarchy commute simultaneously with
the derivatives $D^{+}_1$ and $D^{-}_2$ entering into the $N=(0|2)$ 2DTL
equation \p{f-toda-n=1},
\begin{eqnarray}
[D^{+}_{1}~,~D^{\pm}_{2n}]=[D^{-}_{2}~,~D^{\pm}_{2n}]=0,
\label{algebrasf2}
\end{eqnarray}
while the fermionic flows $D^{+}_{2n+1}$ do not.
Due to this reason the bosonic flows $D^{\pm}_{2n}$
(\ref{flows+F}--\ref{flows-F}) form symmetries of the $N=(0|2)$ 2DTL
equation \p{f-toda-n=1}, while the fermionic flows $D^{+}_{2n+1}$ do not.
Conversely, the $N=(0|2)$ 2DTL equation \p{f-toda-n=1} forms the
infinite-dimensional group of the discrete Darboux-Baeklund symmetries for
the hierarchy of the bosonic flows $D^{\pm}_{2n}$
(\ref{flows+F}--\ref{flows-F}) (particularly, eqs.
(\ref{n=2}--\ref{ds-eq}) ). In other words, if the set
$\{F_{j}~,~{\overline F}_{j}\}$ is a solution of this hierarchy, then the
set $\{F_{j+1}~,~{\overline F}_{j+1}\}$, related to the former by eqs.
\p{f-toda-n=1}, is a solution of the hierarchy as well.

\subsection{Fermionic symmetries of $N=(0|2)$ 2DTL equation}

In this subsection we construct fermionic symmetries of the $N=(0|2)$ 2DTL
equation \p{f-toda-n=1} and their algebra. This construction is similar to
the construction of fermionic symmetries of the STL hierarchy considered in
the subsection 3.3. This permits one to present here its main steps in a
telegraphic style and refer the reader to the subsection 3.3 for more
details.

First, let us consider eqs. \p{equationssf1} at $n=1$,
\begin{eqnarray}
D^{+}_1 u^{(2n)}_{k,j} + (-1)^{k}u^{(2n)}_{k,j}(u_{1,j-k+2n}-u_{1,j})
=u^{(2n)}_{k+1,j+1} +(-1)^{k}u^{(2n)}_{k+1,j}. ~~~
\label{equationss1newf1}
\end{eqnarray}
Substituting
\begin{eqnarray}
u_{1,j-k+2n} - u_{1,j}=
(D^{-}_2)^{-1}(v_{1,j}+v_{1,j+1}-v_{1,j-k+2n}-v_{1,j-k+2n+1}) ~~~
\label{equationss1newf2}
\end{eqnarray}
derived from eqs. \p{recsf1}, into eqs. \p{equationss1newf1} the latter
become
\begin{eqnarray}
&&D^{+}_1 u^{(2n)}_{k,j} + (-1)^{k}u^{(2n)}_{k,j}
(D^{-}_2)^{-1}(v_{1,j}+v_{1,j+1}-v_{1,j-k+2n}-v_{1,j-k+2n+1}) ~~~
\nonumber\\
&&=u^{(2n)}_{k+1,j+1} +(-1)^{k}u^{(2n)}_{k+1,j},
\label{equationss1newf3}
\end{eqnarray}
where $v_{1,j}$ should be expressed in terms of $\{F_j,{\overline F}_j\}$ via
eqs. \p{resol2}. The derived system of equations \p{recsf1} and
\p{equationss1newf3} for the functionals $u^{(2n)}_{k,j}$ can be treated as
the result of the application of the recursive chain of the substitutions
\p{recsf1} to the symmetry equation corresponding to the symmetries
$D^{+}_{2n}$ \p{flows+F} of the $N=(0|2)$ 2DTL equation \p{f-toda-n=1}.
Equivalently, this system represents the consistency condition for the
algebra \p{algebrasf2} realized on the shell of the $N=(0|2)$ 2DTL equation
\p{f-toda-n=1}. Therefore, one can construct the relevant, for the problem
under consideration, solutions of equations \p{recsf1} and
\p{equationss1newf3} forgetting about both the way how they were actually
derived and their relation to the $N=(0|2)$ hierarchy.

It is a matter of simple direct calculations to verify
that eqs. \p{recsf1} and \p{equationss1newf3} possess both bosonic
\begin{eqnarray}
u^{(2n)}_{0,j}=1, \quad u^{(2n)}_{l,j}=0, \quad l<0
\label{solbos}
\end{eqnarray}
and fermionic
\begin{eqnarray}
u^{(2n)}_{-1,j}=(-1)^{j+1}{\epsilon},
\quad u^{(2n)}_{l,j}=0, \quad l < -1
\label{solfer}
\end{eqnarray}
solutions where ${\epsilon}$ is a dimensionless fermionic constant.

The bosonic solution \p{solbos} corresponds to the bosonic symmetries
$D^{+}_{2n}$ \p{flows+F} of the $N=(0|2)$ 2DTL equation \p{f-toda-n=1}
discussed in the previous subsection (see the paragraph after eq.
\p{rec2basiss1f}).

Now, we would like to concentrate on the fermionic solution \p{solfer}
aiming to elaborate the corresponding fermionic symmetries we are looking for.
Let us represent the bosonic time derivative $D^{+}_{2n}$ corresponding to
the solution \p{solfer} and the functionals $u^{(2n)}_{k,j}$ entering into
eqs. \p{flows+F}, \p{recsf1}, \p{solfer} and \p{algebrasf2} in the following
form:
\begin{eqnarray}
D^{+}_{2n} := {\epsilon} {\cal D}^{+}_{2n+1}, \quad
u^{(2n)}_{k,j}:={\epsilon}{\cal U}^{(2n+1)}_{k+1,j}
\label{evolderf}
\end{eqnarray}
defining a new fermionic evolution derivative ${\cal D}^{+}_{2n+1}$ and
functionals ${\cal U}^{(2n+1)}_{k,j}$. Then the fermionic constant
${\epsilon}$ enters linearly into both sides of eqs. \p{flows+F},
\p{recsf1}, \p{solfer} and \p{algebrasf2} which now become
\begin{eqnarray}
{\cal D}^{+}_{2n+1} \ln D^{+}_1 {\overline F}_j &=&
{\cal U}^{(2n+1)}_{2n+1,2j}-{\cal U}^{(2n+1)}_{2n+1,2j-1}, \nonumber\\
\nonumber\\ {\cal D}^{+}_{2n+1} \ln D^{+}_1 F_j &=&
{\cal U}^{(2n+1)}_{2n+1,2j-1}-{\cal U}^{(2n+1)}_{2n+1,2(j-1)},
\label{flows+Fnew1}
\end{eqnarray}
\begin{eqnarray}
D^{-}_{2}{\cal U}^{(2n+1)}_{k,j}&=&
v_{0,j}{\cal U}^{(2n+1)}_{k-2,j-2} -
v_{0,j-k+2n+3}{\cal U}^{(2n+1)}_{k-2,j} \nonumber\\
&-&v_{1,j}{\cal U}^{(2n+1)}_{k-1,j-1} +
(-1)^{k}v_{1,j-k+2n+2}{\cal U}^{(2n+1)}_{k-1,j},
\label{recsf1new}
\end{eqnarray}
\begin{eqnarray}
{\cal U}^{(2n+1)\pm}_{0,j}=(-1)^{j+1}, \quad
{\cal U}^{(2n+1)\pm}_{-1,j}=0
\label{recs1new1f}
\end{eqnarray}
and
\begin{eqnarray}
\{D^{+}_1~,~{\cal D}^{+}_{2n+1}\}=
[D^{-}_2~,~{\cal D}^{+}_{2n+1}]=0,
\label{algebra2newf1}
\end{eqnarray}
respectively. From these equations we see that the fermionic flows
${\cal D}^{+}_{2m+1}$ actually do not depend on ${\epsilon}$ and
anticommute (commute) with the fermionic (bosonic) derivative $D^{+}_1$
($D^{-}_2$) \p{algebra2newf1} entering into the $N=(0|2)$ 2DTL equation
\p{f-toda-n=1}; so ${\cal D}^{\pm}_{2m+1}$ form fermionic symmetries
of the latter.

Now, let us establish the algebra of the fermionic symmetries
${\cal D}^{+}_{2n+1}$ (\ref{flows+Fnew1}--\ref{recs1new1f}).

First using eqs. \p{flows+Fnew1} and \p{algebra2newf1}
we calculate the fermionic symmetry ${\cal D}^{+}_1$
\begin{eqnarray}
{\cal D}^{+}_1 F_j=-D^{+}_1 F_j, \quad {\cal D}^{+}_1 {\overline F}_j
= D^{+}_1 {\overline F}_j
\label{flows+Fanew3}
\end{eqnarray}
and its algebra
\begin{eqnarray}
\{{\cal D}^{+}_{1}~,~{\cal D}^{+}_{1}\}=
-\{D^{+}_{1}~,~D^{+}_{1}\}\equiv -2D^{+}_{2}.
\label{recs1new4f}
\end{eqnarray}
Then, we use eqs. \p{flows+Fanew3} in order to replace $D^{+}_{1}$
by ${\cal D}^{+}_{1}$ in the expressions for both the bosonic
$D^{\pm}_{2n}$ (\ref{flows+F}--\ref{flows-F}) and fermionic
${\cal D}^{+}_{2n+1}$ (\ref{flows+Fnew1}--\ref{recs1new1f}) symmetries
transforming them into the new basis
\begin{eqnarray}
&&{\widehat u}^{(2n+1)}_{k,j}:=
c_k (-1)^{(k+1)(j+1)}{\cal U}^{(2n+1)}_{k,j}, \quad
{\widehat u}^{(2n)}_{k,j}:= c_k (-1)^{kj}u^{(2n)}_{k,j}, \nonumber\\
&&{\widehat v}^{(m)}_{k,j}:= (-1)^{k}{\widetilde v}^{(m)}_{k,j}, \quad
c_{2n}= c_{2n+1}\equiv (-1)^{n},\nonumber\\ &&{\widehat {\cal
D}}^{+}_{2n+1}:= c_{2n+1}{\cal D}^{+}_{2n+1}, \quad
{\widehat {\cal D}}^{+}_{2n}:=
c_{2n}D^{+}_{2n}, \quad {\widehat {\cal D}}^{-}_{2n}:=D^{-}_{2n}
\label{recs1new5ff2}
\end{eqnarray}
which is defined by a single requirement that the form of the symmetries
in this basis is as close as possible to the form of the flows
$D^{+}_{n}$ and $D^{-}_{2n}$ (\ref{flows+F}--\ref{flows-F})
of the $N=(0|2)$ 2DTL hierarchy whose algebra \p{algebrasf1} is known. In the
new basis \p{recs1new5ff2} the symmetries ${\cal D}^{+}_{2n+1}$
(\ref{flows+Fnew1}--\ref{recs1new1f}) and ${\cal D}^{\pm}_{2n}$
(\ref{flows+F}--\ref{flows-F}) as well as the algebra \p{recs1new4f} become
\begin{eqnarray}
{\widehat {\cal D}}^{+}_{n} \ln
{\widehat {\cal D}}^{+}_1 {\overline F}_j &=&
{\widehat u}^{(n)}_{n,2j}-{\widehat u}^{(n)}_{n,2j-1}, \nonumber\\
{\widehat {\cal D}}^{+}_{n} \ln {\widehat {\cal D}}^{+}_1 F_j &=&
{\widehat u}^{(n)}_{n,2j-1}-{\widehat u}^{(n)}_{n,2(j-1)}, \nonumber\\
{\widehat {\cal D}}^{-}_{2}{\widehat u}^{(n)}_{k,j}&=&
v_{0,j}{\widehat u}^{(n)}_{k-2,j-2} -
v_{0,j-k+n+2}{\widehat u}^{(n)}_{k-2,j} \nonumber\\
&+&(-1)^{n}v_{1,j}{\widehat u}^{(n)}_{k-1,j-1} +
(-1)^{k}v_{1,j-k+n+1}{\widehat u}^{(n)}_{k-1,j}, \nonumber\\
{\widehat u}^{(n)\pm}_{0,j}&=&1, \quad
{\widehat u}^{(n)\pm}_{-1,j}=0,
\label{recs1new1fnew}
\end{eqnarray}
and
\begin{eqnarray}
&&{\widehat {\cal D}}^{-}_{2n}\ln {\widehat {\cal D}}^{+}_1{\overline F}_j =
-{\widehat v}^{(n)}_{2n,2j}+{\widehat v}^{(n)}_{2n,2j-1}, \nonumber\\
&&{\widehat {\cal D}}^{-}_{2n}\ln {\widehat {\cal D}}^{+}_1 F_j =
-{\widehat v}^{(n)}_{2n,2j-1}+{\widehat v}^{(n)}_{2n,2(j-1)}, \nonumber\\
&&{\widehat {\cal D}}^{+}_1 {\widehat v}^{(m)}_{k,j}=
g_{j+1}{\widehat v}^{(m)}_{k-1,j+1} +
(-1)^{k}g_{j+k-2m}{\widehat v}^{(m)}_{k-1,j}, \quad
{\widehat v}^{(n)\pm}_{0,j}=1
\label{rec2basiss1fnew}
\end{eqnarray}
as well as
\begin{eqnarray}
\{{\widehat {\cal D}}^{\mp}_{1}~,~
{\widehat {\cal D}}^{\mp}_{1}\}=2{\widehat {\cal D}}^{\mp}_{2},
\label{algebras1newff}
\end{eqnarray}
respectively, where
\begin{eqnarray}
&&g_{2j-1}=-{\widehat {\cal D}}^{+}_1 F_j, \quad
g_{2j}=+{\widehat {\cal D}}^{+}_1 {\overline F}_j, \nonumber\\
&&v_{0,2j+1}=0, \quad \quad \quad
v_{0,2j}=-({\widehat {\cal D}}^{+}_1F_j)
{\widehat {\cal D}}^{+}_1{\overline F}_j, \nonumber\\
&&v_{1,2j} =  F_j{\widehat {\cal D}}^{+}_1 {\overline F}_j, \quad
v_{1,2j-1} = -({\widehat {\cal D}}^{+}_1 F_j) {\overline F}_j.
\label{resol1f}
\end{eqnarray}
The relations (\ref{recs1new1fnew}--\ref{resol1f}) completely reproduce
the corresponding relations (\ref{flows+F}--\ref{flows-F}),
(\ref{recsf1}--\ref{recsf2}), \p{recs1new4f}, \p{resol1} and \p{resol2} for
the flows of the $N=(0|2)$ 2DTL hierarchy where, however, the evolution
derivatives $D^{\pm}_{2n}$ and $D^{+}_{2n+1}$ are replaced by ${\widehat
{\cal D}}^{\pm}_{2n}$ and ${\widehat {\cal D}}^{+}_{2n+1}$, respectively.
Therefore, one can conclude that the algebra of the evolution derivatives
${\widehat {\cal D}}^{\pm}_{2n}$ and ${\widehat {\cal D}}^{+}_{2n+1}$ have
also to reproduce the algebra of the evolution derivatives $D^{\pm}_{2n}$ and
$D^{+}_{2n+1}$ \p{algebrasf1}. Thus, we are led to the following formulae
for both this algebra and the algebras \p{algebrasf2}, \p{algebra2newf1}
as well as \p{recs1new4f} transformed to the basis \p{recs1new5ff2}
\begin{eqnarray}
&&\{D^{+}_{1}~,~D^{-}_{2}\}=0, \quad \{D^{+}_{1}~,~D^{+}_{1}\}=
-2{\widehat {\cal D}}^{+}_{2}, \nonumber\\
&& [D^{+}_1~,~{\widehat {\cal D}}^{\pm}_{2n}]=
\{D^{+}_1~,~{\widehat {\cal D}}^{+}_{2n+1}\}=
[D^{-}_2~,~{\widehat {\cal D}}^{\pm}_{2n}]=
[D^{-}_2~,~{\widehat {\cal D}}^{+}_{2n+1}]=0,\nonumber\\
&& [{\widehat {\cal D}}^{+}_{n}~,~
{\widehat {\cal D}}^{\pm}_{2l}]=
[{\widehat {\cal D}}^{-}_{2n}~,~{\widehat {\cal D}}^{-}_{2l}]=0,\quad
\{{\widehat {\cal D}}^{+}_{2n+1}~,~
{\widehat {\cal D}}^{+}_{2l+1}\}=2{\widehat {\cal D}}^{+}_{2(n+l+1)}.~~~~~~
\label{algebras1new1f}
\end{eqnarray}
The symmetries ${\cal D}^{+}_{2n+1}$ (\ref{flows+Fnew1}--\ref{recs1new1f})
of the $N=(2|2)$ 2DTL equation \p{f-toda-n=1} are actually also symmetries
of all the bosonic flows $D^{\pm}_{2l}$ (\ref{flows+F}--\ref{flows-F}) of
the $N=(2|2)$ 2DTL hierarchy because of the commutation relations
\begin{eqnarray}
[{\cal D}^{+}_{2n+1}~,~D^{\pm}_{2l}]=0
\label{algebras1new2symf}
\end{eqnarray}
following from the algebra \p{algebras1new1f} and relations
\p{recs1new5ff2}.

The existence of the fermionic symmetries ${\cal D}^{+}_{2n+1}$
(\ref{flows+Fnew1}--\ref{recs1new1f}) creates a new interesting problem of
constructing both additional evolution equations for the Lax operators
$L^{\pm}$ \p{laxsf1} generated by ${\cal D}^{+}_{2n+1}$ and
commutation relations between the latter and the fermionic flows
$D^{+}_{2n+1}$ \p{flows+F} of the $N=(0|2)$ 2DTL hierarchy. We hope
to return to this problem in future.

To close this section, let us point out that the $N=(0|2)$ supersymmetry
and $N=(0|2)$ superfield formulation of both the $N=(0|2)$ 2DTL equation
\p{f-toda-n=1} and the bosonic flows $D^{\pm}_{2l}$
(\ref{flows+F}--\ref{flows-F}) of the $N=(0|2)$ 2DTL hierarchy can easily be
uncovered from the approach and formulae of this subsection. In order to
see that, it is enough only to introduce a new, $N=2$ basis
$\{D_{+},~{\overline D}_{+}\}$ in the space of the fermionic evolution
derivatives $\{D^{+}_{1},~{\cal D}^{+}_{1}\}$, namely:
\begin{eqnarray}
D_{+}:=\frac{1}{2}({\cal D}^{+}_1 + D^{+}_1), \quad
{\overline D}_{+}:=\frac{1}{2}( D^{+}_1-{\cal D}^{+}_1)
\label{sup1}
\end{eqnarray}
which form the algebra of the $N=2$ supersymmetry
\begin{eqnarray}
D^2_{+}={\overline D}^2_{+}=0, \quad
\{D_{+}~,~{\overline D}_{+}\}={\partial}_{+},
\label{sup1alg}
\end{eqnarray}
where we have introduced the notation
$(D^{\pm}_1)^2\equiv {\partial}^{\pm}_2:={\partial}_{\pm}$. Then,
relations \p{flows+Fanew3} become
\begin{eqnarray}
D_{+} F_j=0,  \quad {\overline D}_{+} {\overline F}_j=0
\label{sup2}
\end{eqnarray}
and have the form of the $N=(0|2)$ chirality constraints for the chiral and
antichiral $N=(0|2)$ superfields $F_j$ and ${\overline F}_j$, respectively.
For illustration, we present the $N=(0|2)$ 2DTL equation \p{f-toda-n=1n}
and the supersymmetric generalization of the Davey-Stewartson equation
\p{ds-eq} identically rewritten to this basis
\begin{eqnarray}
{\partial}_{-} \ln \Bigl(({\overline D}_{+}F_{j+1})(D_{+}{\overline F})\Bigr)
= -F_j{\overline F}_j + F_{j+1}{\overline F}_{j+1}
\label{f-toda-n=1n-n2}
\end{eqnarray}
and
\begin{eqnarray}
D^{+}_4 F_{j}&=&-{\partial}_{+}^2F_{j}+
2D_{+}\Bigl(({\overline D}_{+} F_{j}){\partial}_{-}^{-1}{\partial}_{+}
(F_j{\overline F}_j)\Bigr), \nonumber\\
D^{+}_4 {\overline F}_j&=&+{\partial}_{+}^2{\overline F}_j+
2{\overline D}_{+}\Bigl((D_{+} {\overline F}_j)
{\partial}_{-}^{-1}{\partial}_{+} (F_j{\overline F}_j)\Bigr),
\label{ds-eq-n2}
\end{eqnarray}
respectively which possess both manifest $N=(0|2)$ supersymmetry and
$N=(0|2)$ superfield form.

\section{Generalizations}
In this section, we briefly describe generalizations of the supersymmetric
$N=(0|2)$ 2DTL hierarchy discussed in the preceding section.

We propose the following set of the consistent operator equations:
\begin{eqnarray}
D^{+}_n L^{+}&=&
(-1)^{n}(((L^{+})^{n}_{*})_{+})^{*} L^{+}
-(L^{+})^{*(n)}((L^{+})^{n}_{*})_{+} \nonumber\\
&+&(1-(-1)^n)(L^{+})^{n+1}_{*}, \nonumber\\
D^{+}_n L^{-}&=&
((L^{+})^{n}_{*})_{+} L^{-}-
(L^{-})^{*(n)}((L^{+})^{n}_{*})_{+}, \nonumber\\
D^{-}_{2Mn} L^{+}&=&
(((L^{-})^{n})_{-})^{*} L^{+}
-(L^{+})((L^{-})^{n})_{-}, \nonumber\\
D^{-}_{2Mn} L^{-}&=&[((L^{-})^{n})_{-}, L^{-}], \quad n \in {\bN},
\label{laxreprsf1clos}
\end{eqnarray}
\begin{eqnarray}
L^{+}=\sum^{\infty}_{k=0} u_{k,j}e^{(1-k){\partial}}, \quad
L^{-}=\sum^{\infty}_{k=0} v_{k,j}e^{(k-2M){\partial}},
\label{laxsf1clos}
\end{eqnarray}
\begin{eqnarray}
u_{0,j}\equiv 1, \quad v_{0,2Mj+1}\equiv 0, \quad v_{0,2Mj+\alpha}\neq 0,
\quad \alpha = 2,3,\ldots, 2M
\label{bound5}
\end{eqnarray}
generating the non-abelian algebra of the flows
\begin{eqnarray}
&&[D^{+}_{n}~,~D^{-}_{2Ml}]=[D^{+}_{2n}~,~D^{+}_{2l}]=
[D^{-}_{2Mn}~,~D^{-}_{2Ml}]=0,\nonumber\\
&&\{D^{+}_{2n+1}~,~D^{+}_{2l+1}\}=2D^{+}_{2(n+l+1)}
\label{algebrasf1clos}
\end{eqnarray}
which may be realized in the superspace  $\{t^{+}_n, t^{-}_{2Mn}\}$
\begin{eqnarray}
D^{\pm}_{2Mn} ={\partial}^{\pm}_{2Mn}, \quad
D^{+}_{2n+1} ={\partial}^{+}_{2n+1}+
\sum^{\infty}_{l=1}t^{+}_{2l-1}{\partial}^{+}_{2(k+l)},
\label{covderfclos}
\end{eqnarray}
where $M\in {\bN}$ is a fixed number and $t^{+}_{2n}~,t^{-}_{2Mn}~$
($t^{+}_{2n+1}$) are bosonic (fermionic) evolution times.
At $M=1$ eqs. (\ref{laxreprsf1clos}--\ref{covderfclos}) reproduce the Lax
pair representation of the $N=(0|2)$ 2DTL hierarchy.

At different values $M$ equations
(\ref{laxreprsf1clos}--\ref{covderfclos}) generate non-equivalent
supersymmetric hierarchies. Nevertheless, any hierarchy with $M=M_1$
can be produced by reduction of the hierarchy with
$M=nM_1$ ($n\in {\bN}$) if the latter is provided by
the following additional reduction constraints: $v_{0,2M_1j+1}\equiv 0$
which are obviously consistent with the original
constraints $v_{0,2nM_1j+1}\equiv 0$ entering into the definition
\p{bound5} of the latter hierarchy.

A detailed analysis of the generalizations proposed here is under way.

\section{Conclusion}

In this paper, we have clarified the origin of fermionic and bosonic solutions
\cite{ly1,ls1,ols1} to the symmetry equations corresponding to the 2DTL and
$N=(2|2)$ supersymmetric 2DTL equations and established the algebras of the
corresponding symmetries. As a byproduct we have also proved the conjecture,
proposed in \cite{ols2}, regarding an $N=(2|2)$ superfield formulation of the
STL hierarchy. Then, we have proposed the new, $N=(0|2)$ supersymmetric 2DTL
hierarchy. Furthermore, we have constructed bosonic and fermionic symmetries
of the $N=(0|2)$ 2DTL equation belonging the hierarchy and their algebra 
to our knowledge for the first time. We have also discussed an $N=(0|2)$
superfield formulation of the $N=(0|2)$ 2DTL hierarchy. Finally, we have
generalized the approach developed for the case of the $N=(0|2)$ 2DTL
hierarchy and proposed an infinite class of new supersymmetric Toda type
hierarchies.

{}~

\noindent{\bf Acknowledgments.}
This work was partially supported by DFG Grant No. 436 RUS 113/359/0
(R), RFBR-DFG Grant No. 99-02-04022, PICS Project No. 593,
RFBR-CNRS Grant No. 98-02-22034, Nato Grant No. PST.CLG 974874,
RFBR Grant No. 99-02-18417.

\end{document}